\documentclass[lettersize,journal]{IEEEtran}
\usepackage[siunitx]{circuitikz}
\usepackage{caption}
\usepackage{xcolor}
\usepackage{amsmath,amsfonts}
\usepackage{amssymb}

\usepackage{algorithmic}
\usepackage{algorithm}
\usepackage{array}
\usepackage{svg}

\usepackage[caption=false,font=normalsize]{subfig}
\usepackage{textcomp}
\usepackage{stfloats}
\usepackage{url}
\usepackage{verbatim}
\usepackage{graphicx}
\usepackage{cite}
\begin{document}

\title{Are Stacked Intelligent Metasurfaces (SIMs) Better than Single-layer Reconfigurable Intelligent Surfaces (RISs) for Wideband Multi-user MIMO Communication Systems?}

\author{\IEEEauthorblockN{
Muhammad Ibrahim, Amine Mezghani, and Ekram Hossain, {\em FIEEE}\thanks{The authors are with the Department of Electrical and Computer Engineering at the University of Manitoba, MB, Canada (emails: ibrah101@myumanitoba.ca, \{Amine.Mezghani, Ekram.Hossain\}@umanitoba.ca).}
}}


\maketitle

\begin{abstract}
Cascaded or stacked intelligent metasurfaces (SIMs) have emerged as a promising technology to overcome the physical limitations of single-layer reconfigurable intelligent surfaces (RISs) in wideband wireless communication. By intelligently manipulating electromagnetic waves, SIMs enhance signal propagation in complex environments and offer additional degrees of freedom for beamforming. {This paper proposes a coupling-aware, wideband, circuit-based framework that captures frequency-dependent mutual coupling and wideband channel responses over multiple subbands. Based on this model}, we formulate a joint active and passive beamforming design that optimizes the base-station precoder to enable carrier aggregation across frequency-selective subbands, together with metasurface phase shifts, to maximize spectral efficiency. Simulation results reveal the importance of accounting for coupling and wideband effects, and show that performance depends strongly on operating conditions. Single-layer RIS configurations can be favorable in narrowband and/or low-SNR regimes, whereas SIMs can significantly outperform under wideband multi-user conditions by mitigating coupling-induced distortion and maintaining a more consistent phase response across frequencies. The results provide physical insights into design trade-offs between structural simplicity and wideband adaptability, highlighting SIMs as a scalable solution for future-generation wideband multi-user MIMO systems. {We further show that partially reconfigurable SIM architectures achieve near-optimal performance with reduced complexity.}

\end{abstract}


\section{Introduction}

\IEEEPARstart{}{}The rapid evolution of 6G networks demands high spectral efficiency, wider bandwidth operation, and energy-efficient hardware architectures. While massive Multiple-Input Multiple-Output (MIMO) offers significant spatial multiplexing gains, its hardware cost and RF chain complexity make it less scalable at mmWave and sub-THz frequencies~\cite{9,10}. This motivates the exploration of energy-efficient and less complex architectures.

Reconfigurable Intelligent Surfaces (RISs) have emerged as a promising solution for energy-efficient wireless communication. By intelligently tuning the electromagnetic response of the impinging signals, RISs can enhance coverage and improve spectral efficiency with minimal power consumption~\cite{28,29,30,31,32,33}.  {Both diagonal RIS (D-RIS) architectures with phase-only control and beyond-diagonal RIS (BD-RIS) architectures~\cite{35} with joint phase and amplitude control have been extensively studied in MIMO and OFDM systems}, investigating aspects such as joint active and passive beamforming, channel estimation, and hardware constraints. However, conventional single-layer RIS architectures remain limited in their ability to support complex signal manipulation and system performance enhancement.

A new technology has recently been introduced in the form of cascaded layers of transmissive RIS, also known as stacked intelligent metasurfaces (SIMs)~\cite{11}. Unlike conventional digital or hybrid beamforming systems, SIMs process electromagnetic (EM) waves directly in the analog domain~\cite{12}, enabling more energy-efficient control of signal propagation. Each layer of a SIM acts as a programmable surface composed of tunable meta-atoms that can adjust the phase of the signal. {This architecture provides additional spatial degrees of freedom through cascaded electromagnetic processing across multiple transmissive layers, enabling more flexible and precise beamforming control}~\cite{17}.

Several recent studies have explored SIMs to enhance spectral efficiency in various communication systems. In~\cite{21}, a SIM-aided integrated sensing and communication system was proposed, jointly optimizing the SIM phase shifts and BS power to mitigate inter-user interference. The work in~\cite{22} extended SIM applications to holographic MIMO systems, formulating joint optimization of phase shifts and transmit covariance matrices. Other studies have investigated channel estimation and direction-of-arrival estimation~\cite{13,15}. However, most existing SIM works rely on narrowband assumptions and overlook frequency dependent phenomena such as beam squint, mutual coupling, and inter-layer reflections, which significantly impact wideband performance. Although~\cite{16} introduced a fully analog wideband SIM beamforming method, it did not incorporate coupling or reflection effects. Moreover, prior SIM research has primarily focused on spatial-domain optimization, while efficient exploitation of frequency-domain characteristics remains largely unexplored.

To accurately capture mutual coupling and mutual reflections in SIMs, multiport network theory has been widely adopted as a physically consistent modeling framework. For instance,~\cite{19} employed cascaded scattering ($S$)-parameter representations to explicitly model wave propagation, inter-element coupling, and inter-layer reflections. {While physically accurate, cascaded $S$-parameter models are often analytically intractable and computationally demanding. To address this issue,~\cite{20} introduced a transmission ($T$)-parameter-based formulation that is mathematically equivalent to the cascaded $S$-parameter model but offers improved tractability}. In parallel,~\cite{3} developed a general impedance ($Z$)-parameter-based SIM model that exploits structured impedance matrices to efficiently capture coupling effects. Nevertheless, these existing multiport SIM models are predominantly developed under narrowband assumptions and do not explicitly account for frequency-dependent impedance variations and coupling effects that arise in wideband systems.

Carrier aggregation (CA), a key technique in 5G networks~\cite{23,24}, enables wideband transmission by combining multiple subband carriers to enhance throughput and spectral efficiency.  {Extending CA to metasurface-assisted systems introduces a joint spatial and frequency-domain design problem, since a common metasurface configuration must operate over multiple subbands. Under wideband operation, frequency-dependent impedance variations and mutual coupling make the effective phase response subband dependent, which can lead to beam squint and phase misalignment across the aggregated carriers. These effects motivate coupling-aware wideband modeling and optimization to maintain consistent multi-subband performance.}

{While metasurface-assisted system modeling and carrier-aggregation-aware transmission design have been studied under simplified assumptions, their joint design under physically consistent, coupling-aware wideband metasurface models remains limited. Moreover, existing SIM studies largely emphasize spatial-domain optimization and often adopt narrowband abstractions that do not capture the frequency-dependent impedance and coupling behavior critical in wideband settings. This motivates a tractable circuit-theoretic framework that captures wideband coupling effects and enables joint optimization of metasurface phase shifts and multi-subband resource allocation.}

To this end, we develop a circuit-theory-based wideband SIM framework that incorporates mutual coupling and inter-layer reflections while enabling joint optimization of phase shifts and resource allocation across subbands. Our main contributions are summarized as follows:
\begin{itemize}
\item Development of a circuit-theoretic model for a SIM-assisted downlink multi-user wideband MIMO system that captures mutual coupling, inter-layer reflections, and frequency selectivity across subbands.
\item Joint optimization of SIM phase shifts and beamforming across users and subbands, improving achievable sum rate relative to narrowband baselines.
\item A partial reconfiguration strategy for SIM that reduces computational and control overhead while maintaining high goodput under practical constraints.
\item A comparative analysis between single-layer RIS and SIM across different user and bandwidth scenarios, identifying operational regimes where SIM provides clear gains.
\end{itemize}

\noindent
\textbf{Notations:} For any matrix $\mathbf{A}$, $\mathbf{A}^T$, $\mathbf{A}^*$, and $\mathbf{A}^H$ represent its transpose, conjugate, and conjugate transpose, respectively. The trace of $\mathbf{A}$ is denoted by $\mathrm{Tr}(\mathbf{A})$. The Kronecker product of two vectors or matrices is represented by $\otimes$. $\mathrm{Diag}(\mathbf{a})$ denotes a diagonal matrix with the elements of vector $\mathbf{a}$ along its diagonal, while $\mathrm{diag}(\mathbf{A})$ returns a vector containing the diagonal elements of the matrix $\mathbf{A}$.
\section{System Model and Assumptions}

\subsection{SIM-Assisted Multi-user Downlink  MIMO Model}
\begin{figure}
    \centering
    \includegraphics[width=1.0\linewidth]{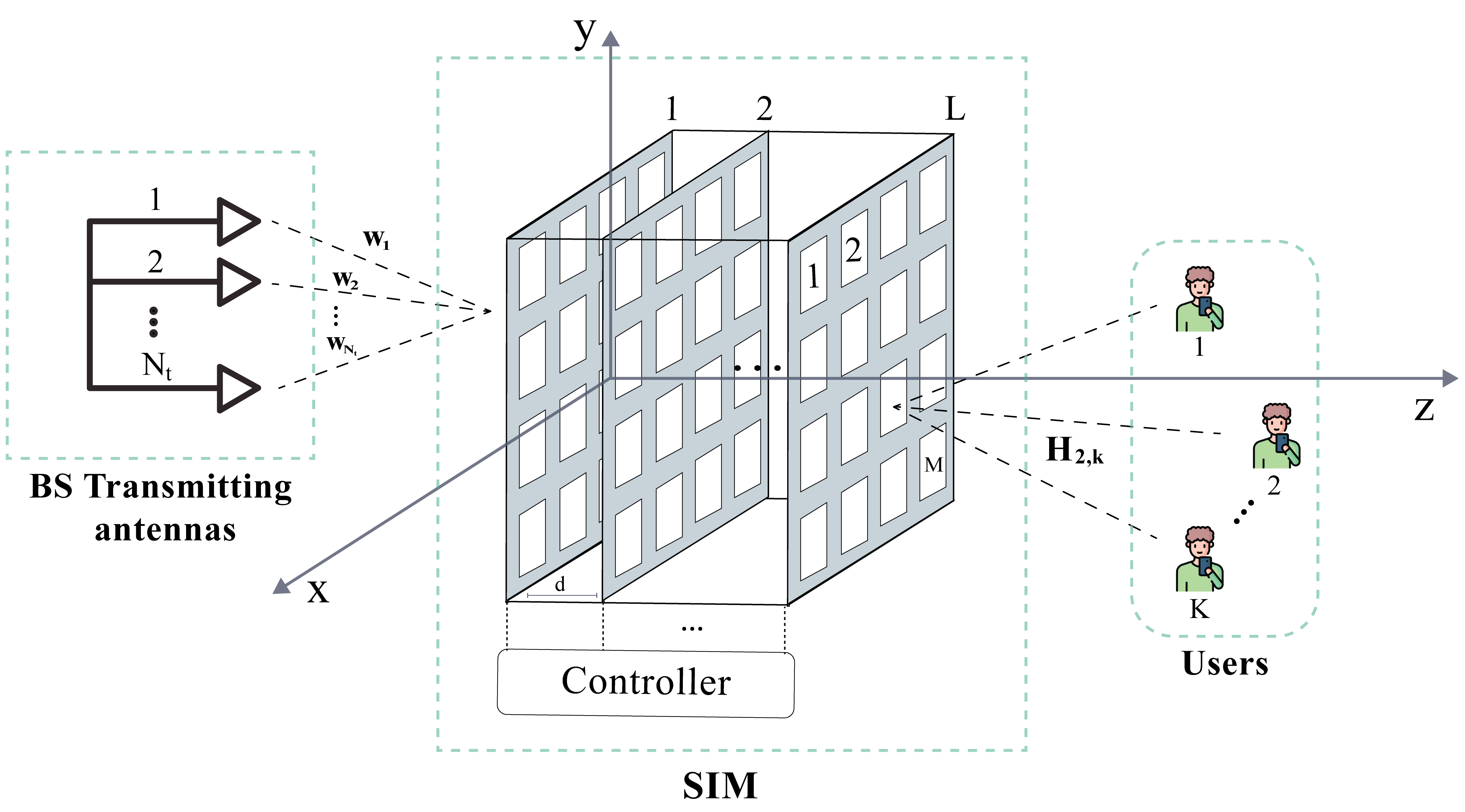}
    \caption{SIM-assisted downlink multi-user MIMO system}
    \label{sys}
\end{figure}

We consider a SIM-assisted multi-user downlink MIMO system, as shown in Fig.~\ref{sys},  {where a SIM is deployed near the BS to assist communication with multiple users}. The BS is equipped with $N_t$ antennas arranged in a uniform linear array (ULA). The SIM consists of $L$ cascaded transmissive metasurface layers with $M$ meta-atoms (or antenna elements) which can be reconfigured by an intelligent controller. The system serves $K$ single-antenna users. {Throughout this paper, the direct BS--UE link is assumed to be blocked, and communication occurs exclusively through the metasurface-assisted path.}

{To enable a fair comparison between single-layer (RIS) and multi-layer (SIM) architectures, the total number of metasurface elements is fixed and denoted by $M_{\mathrm{total}}$. In the single-layer RIS case, all $M_{\mathrm{total}} = M_x \times M_y$ elements are placed on a single planar surface. In the multi-layer case, the same total number of elements is evenly distributed across $L$ cascaded layers, such that each layer contains $M_{\mathrm{total}}/L = M_x \times (M_y/L)$ meta-atoms arranged on a planar grid with fixed inter-element spacing. Interactions are assumed to occur only between adjacent metasurface layers, consistent with the cascaded SIM structure.}

{
The system operates over a wideband channel with center frequency $f_c$ and total bandwidth $B$. The wideband channel is modeled using a multi-carrier framework, where the total bandwidth $B$ is divided into $N_f$ subbands with center frequencies $f_n$, $n = 1, \ldots, N_f$, and the channel response within each subband is assumed to be approximately frequency-flat. Carrier aggregation is modeled in an abstract manner as wideband power allocation. Digital baseband processing is done at the BS across subbands, whereas the SIM is modeled as a passive analog multiport network whose frequency-dependent response is characterized by impedance parameters. The metasurface phase shifts and the transmit power across subbands are jointly optimized to maximize the achievable sum rate of all users. The metasurface phase shifts are assumed to be common across all subbands, while the effective electromagnetic response varies with frequency due to frequency-dependent self-impedance and mutual coupling. These frequency-dependent effects are explicitly captured through a proposed circuit-theoretic modeling framework (as will be presented in the next subsection) and play a key role in characterizing wideband performance differences between single-layer RIS and multi-layer SIM architectures.
}

\subsection{Circuit-Theoretic Model}

The circuit-theoretic model provides a clear and physically consistent foundation for analyzing a MIMO wireless communication model \cite{2}. In this approach, antennas are represented as ports in a multiport network, where transmitted and received signals are described by the corresponding port voltages and currents. Capturing the relationship among these port variables is essential for accurately characterizing the communication model. 

\begin{figure*}[t]
  \centering
  \includegraphics[width=0.75\textwidth]{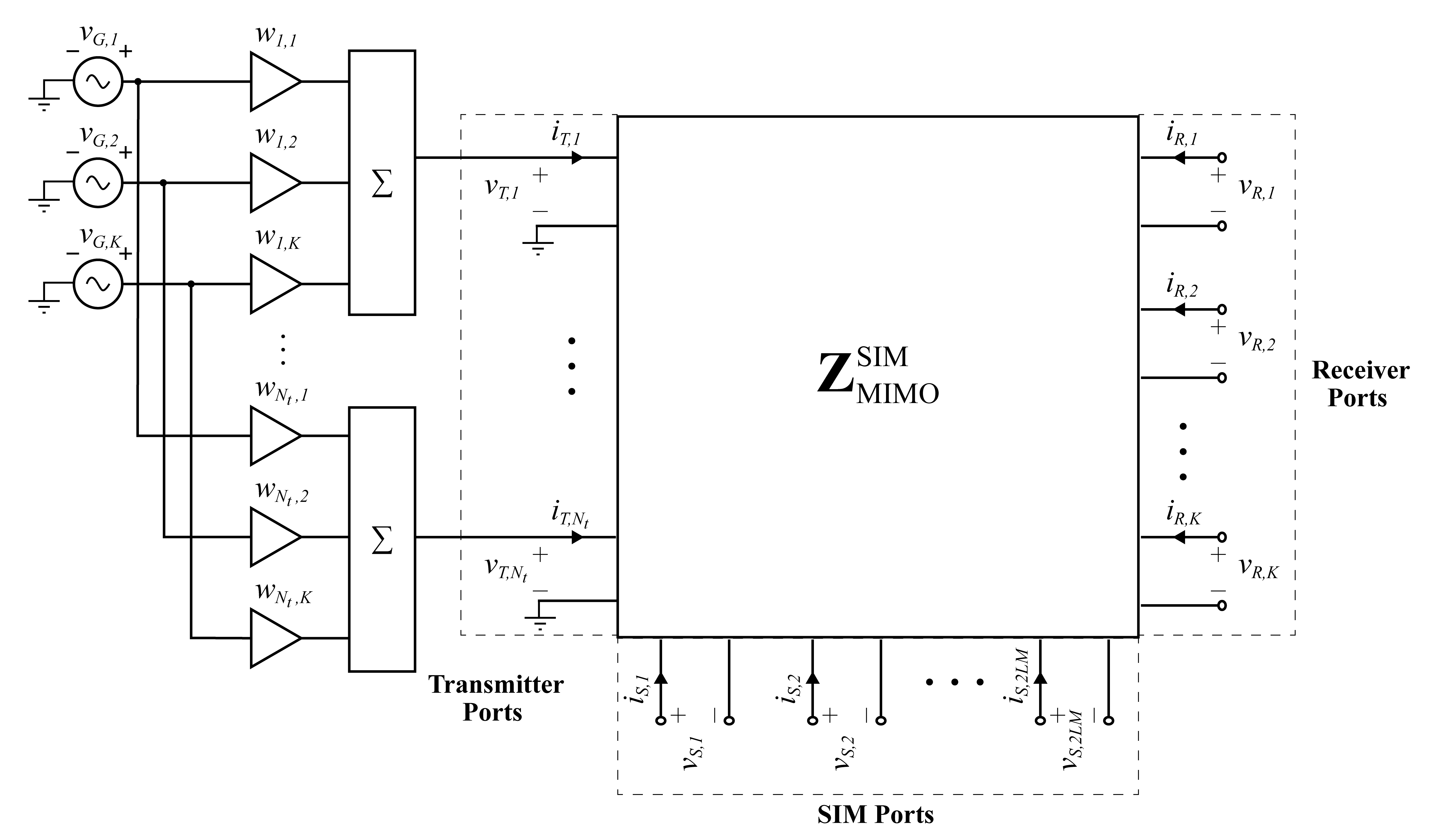}

  \caption{Equivalent circuit-theoretic model for SIM-assisted MIMO communication system}
  \label{fig:2}
\end{figure*}

As shown in Fig.~\ref{fig:2}, the SIM-assisted MIMO communication channel is modeled using multiport network theory with a total of $(N_t + 2M_{\mathrm{total}} + K)$ ports, corresponding to $N_t$ transmit antenna ports, $2M_{\mathrm{total}}$ SIM ports arising from the two facing sides of each transmissive metasurface element, and $K$ single antenna user ports. The Ohm's law governs the relationship between the port voltages and currents given as:
\begin{equation}
    \bold{v}(f) = \bold{Z}^{SIM}_{MIMO}(f) \bold{i}(f),
\end{equation}
where $\bold{Z}^{SIM}_{MIMO}(f)$ is the global impedance matrix as shown in Fig. \ref{fig:2}. This matrix fully characterizes both the antenna subsystems and the wireless propagation channel. 

{To define port variables explicitly, we denote by $\mathbf{v}_T(f), \mathbf{i}_T(f) \in \mathbb{C}^{N_t\times 1}$, $\mathbf{v}_S(f), \mathbf{i}_S(f) \in \mathbb{C}^{2M_{\mathrm{total}}\times 1}$, and $\mathbf{v}_R(f), \mathbf{i}_R(f) \in \mathbb{C}^{K\times 1}$ the port voltage and current vectors at the transmitter (T), SIM (S), and receiver (R), respectively. For the SIM, $\mathbf{v}_S(f)$ and $\mathbf{i}_S(f)$ include the port voltages and currents of all metasurface elements across all cascaded layers, where $2M_{\mathrm{total}}$ denotes the total number of metasurface ports in the SIM.} In particular, we consider $K$ independent generator voltages represented by $\mathbf{v}_G \in \mathbb{C}^{K\times 1}$. These generator voltages are amplified through a gain matrix $\mathbf{W}$, similar to digital beamforming, producing the effective excitation voltages $\mathbf{v}_T$ which serve as inputs to the transmit antenna ports. At the receiver, the load voltages are given by $\mathbf{v}_R$ for $K$ users with matched load impedances collected in $\mathbf{R}=\mathrm{diag}(R_1,\dots,R_K)$.

In this framework, the SIM (S) is treated as a general group of $2M_{\mathrm{total}}$ ports. The corresponding SIM related submatrix of $\mathbf{Z}(f)$ captures the self-impedance of each element as well as its coupling with neighboring elements, the transmitter, and the receiver. 
Hence, the full block structure of the impedance matrix can be expressed as:
\begin{equation}
\begin{bmatrix}
\mathbf{v}_T \\
\mathbf{v}_S \\
\mathbf{v}_R
\end{bmatrix}
=
\begin{bmatrix}
\mathbf{Z}_{TT} & \mathbf{Z}_{TS} & \mathbf{Z}_{TR} \\
\mathbf{Z}_{ST} & \mathbf{Z}_{SS} & \mathbf{Z}_{SR} \\
\mathbf{Z}_{RT} & \mathbf{Z}_{RS} & \mathbf{Z}_{RR}
\end{bmatrix}
\begin{bmatrix}
\mathbf{i}_T \\
\mathbf{i}_S \\
\mathbf{i}_R
\end{bmatrix}.
\label{eq:multiport}
\end{equation}

The voltages and currents at the SIM are related through $\mathbf{v}_S = -\mathbf{Z}_S \mathbf{i}_S, $where $\mathbf{Z}_S$ denotes the impedance matrix associated with the network 
to which the SIM ports are connected. Moreover, in Eq. (\ref{eq:multiport}), the diagonal blocks $\mathbf{Z}_{TT}$, $\mathbf{Z}_{SS}$, and $\mathbf{Z}_{RR}$ represent the internal impedance structure of the transmit array, the SIM, and the receive array, respectively. The diagonal entries within each block correspond to the self-impedances of the ports, while the off-diagonal entries within each block quantify the intra-array coupling effects between different antenna elements.

The off-diagonal blocks, such as $\mathbf{Z}_{TS}$, $\mathbf{Z}_{TR}$ or $\mathbf{Z}_{SR}$, describe the cross-interactions between different subsystems and are therefore referred to as transimpedance matrices. More specifically, for $X, X' \in \{T, S, R\}$ with $X \neq X'$, the matrix $\mathbf{Z}_{XX'} \in \mathbb{C}^{N_X \times N_{X'}}$ characterizes the coupling between ports of subsystem $X$ and ports of subsystem $X'$. Using the electromagnetic reciprocity, these blocks satisfy $\mathbf{Z}_{XX'} = \mathbf{Z}_{X'X}^{\top}$.

The input-output relationship between the transmit and receive voltages can be expressed as 
$\mathbf{v}_R = \mathbf{H}_Z \mathbf{v}_T $, where $\mathbf{H}_Z$ denotes the transfer function. To study the effects of the SIM, we assume that both the transmit and receive antennas are terminated at the reference impedance ($Z_0 = 50$) and that there is no mutual coupling among them. Furthermore, we neglect mutual reflections between the transmitter and the SIM, as well as between the SIM and the receiver. Under these assumptions, and following \cite{4}, the transfer function is given by
\begin{equation}
\mathbf{H}_Z = \frac{1}{4 Z_0} 
\Big[ \mathbf{Z}_{RT} - 
\mathbf{Z}_{RS} \big(\mathbf{Z}_{SS} + \mathbf{Z}_S\big)^{-1} 
\mathbf{Z}_{ST} \Big],
\label{eq:transfer}
\end{equation}
where $Z_0$ is the reference impedance and $Z_S$ defines the controllable load network that is to be optimized. 
\subsection{Model for Mutual Coupling Among SIM Elements}
\label{sec2:B}
\begin{figure}[t]
  \centering
  \begin{circuitikz}[american, thick, line cap=round, line join=round,
                     scale=0.9, transform shape] 
    \coordinate (BL) at (0,0);
    \coordinate (TL) at (0,3);
    \coordinate (TR) at (7,3);
    \coordinate (BR) at (7,0);
    \coordinate (MT) at (3.5,3);
    \coordinate (MB) at (3.5,0);

    \draw (BL) to[open, o-o] (TL);
    \node[left] at (TL) {$+$};
    \node[left] at (BL) {$-$};
    \node[left] at (-0.55,1.5) {$v(f)$};

    \draw[->] (0.25,3) -- (1.15,3) node[midway,below=2pt] {$i(f)$};
    \draw (TL) -- (1.6,3)
          to[C, l^={$C_{ant}=\frac{a}{cR}$}] (2.8,3)
          -- (TR);

    \draw (MB) to[L, l_={$L_{ant}=\frac{aR}{c}$}] (MT);

    \draw (TR) to[R, l={$R$}] (BR);

    \draw (BR) -- (BL);

    \path (BL) -- (MB)
      coordinate[pos=0.38] (zL)
      coordinate[pos=0.82] (zR);

    \draw[->, >=Latex, line cap=round]
      ([yshift=10mm]zL) -- ([yshift=10mm]zR) 
      node[midway, above=1pt] {$Z(f)$};
  \end{circuitikz}

  \caption{Equivalent circuit model (TM$_1$ antenna) }
  \label{fig:sim-circuit}
\end{figure}
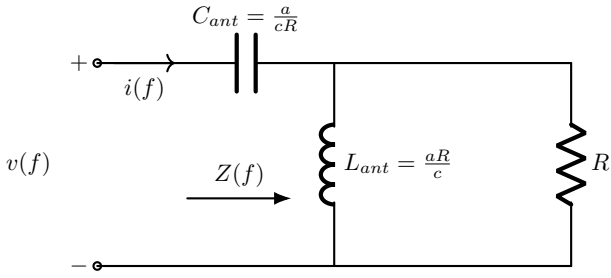

We model all antenna elements as canonical minimum scattering (CMS) antennas as given in \cite{1}. It is well established that maximum bandwidth (lowest $Q$ factor) is obtained when an antenna operates is the lowest order scattering mode. Therefore, we restrict our analysis to CMS antennas operating in transverse-magnetic-1 (TM$_1$) mode only, as the configuration ensures the broadest possible bandwidth. 
The larger amount of energy is stored (non-radiated) in the sphere enclosing the antenna if the radiation mode is higher. Notably, many common and simple antenna structures, such as dipoles, can be effectively approximated as TM$_1$ antennas. Their behavior can be represented through the equivalent circuit model as in Fig. \ref{fig:sim-circuit}\cite{1}. The input voltage and current of the equivalent circuit is given by:
\vspace{-2pt}
\begin{subequations}\label{eq:vi}
\begin{align}
v(f) &= 
\sqrt{\frac{8 \pi \eta}{3}} 
\frac{\sqrt{R} A_{1}}{k} 
\left( 1 + \frac{1}{jka} - \frac{1}{j(ka)^{2}} \right) 
e^{-jka}, \label{eq:vf}\\
i(f) &=
- \sqrt{\frac{8 \pi \eta}{3}} 
\frac{A_{1}}{\sqrt{R} k} 
\left( 1 + \frac{1}{jka} \right) 
e^{-jka}, \label{eq:if}
\end{align}
\end{subequations}
where $A_1$ denotes the complex coefficient of the TM$_1$ mode (Fig.~\ref{fig:sim-circuit}), 
$k = \tfrac{2\pi f}{c}$ is the wavenumber, and 
$\eta = \sqrt{\tfrac{\mu}{\epsilon}}$ is the intrinsic impedance of the medium, 
with $\mu$ and $\epsilon$ representing its permeability and permittivity, respectively. 
Furthermore, $R$ is the radiation resistance of the antenna, and $a$ is the radius of the antenna enclosed in a sphere \cite{1}. 
Additionally, $L_{ant}$, $C_{ant}$, and $R$ denote the equivalent inductance, capacitance, and resistance in the TM$_1$ circuit representation of the antenna.
By applying basic circuit analysis, the corresponding self-impedance can be obtained as:
\vspace{-5pt}
\begin{equation}
Z(f) = \frac{v(f)}{i(f)} 
= \left[ \frac{c^{2} + j 2 \pi f c a - (2 \pi f a)^{2}}
{j 2 \pi f c a - (2 \pi f a)^{2}} \right] R.
\label{self}
\end{equation}

\begin{figure}
    \centering
    \includegraphics[width=0.7\linewidth]{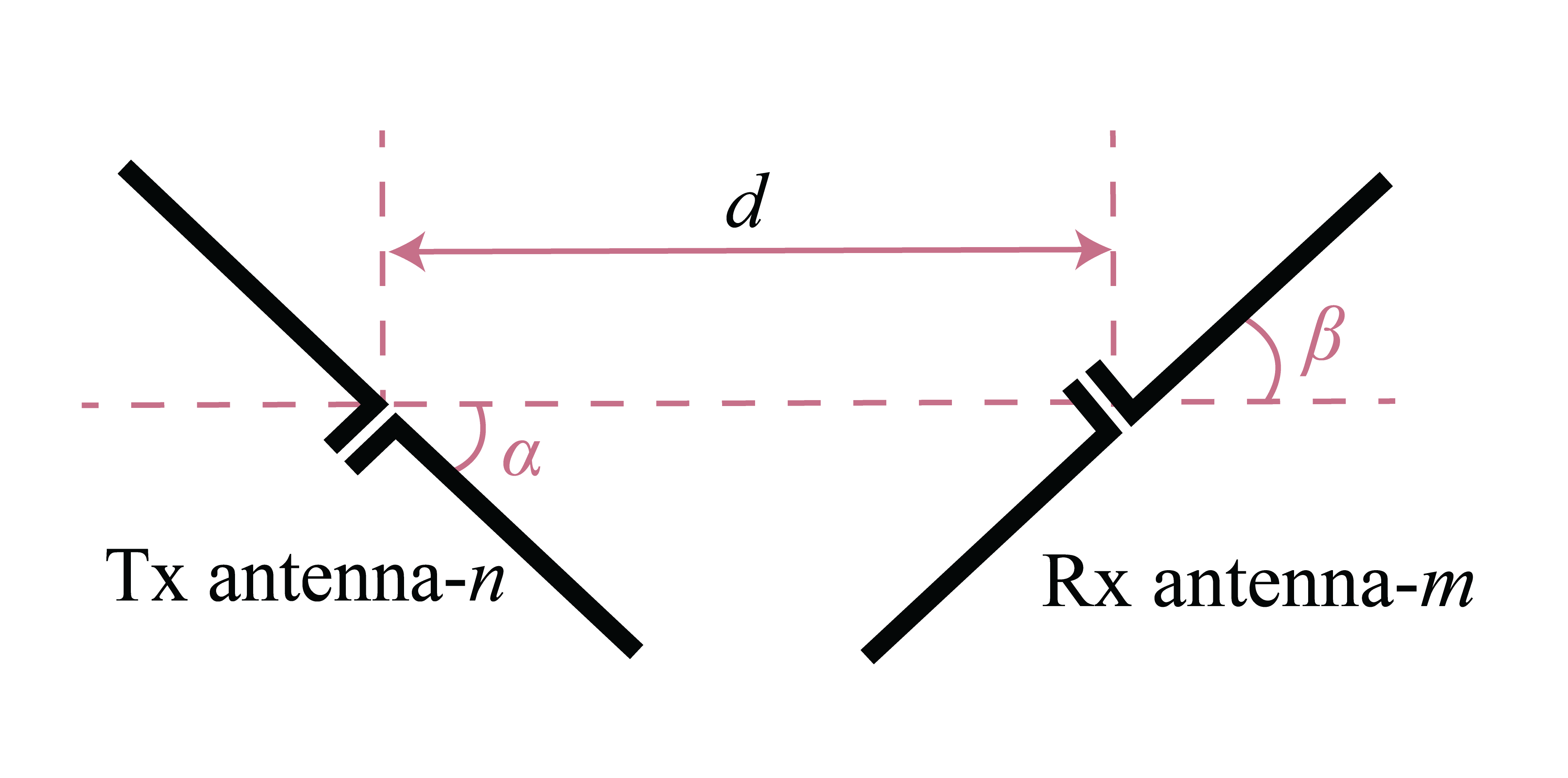 }
    \caption{Two $TM_1$ antennas arbitrarily oriented and seperated by a distance $d$}
    \label{fig:3}
\end{figure}

Using \eqref{self}, the diagonal entries of $\mathbf{Z}_{TT}$ and $\mathbf{Z}_{RR}$ can be directly obtained. 
In addition, the mutual impedance between two TM$_1$ antennas indexed by $n$ and $m$, 
separated by a distance $d$ and oriented with angles $\alpha$ and $\beta$ along the line joining both antennas, as illustrated in Fig.~\ref{fig:3}, 
is expressed as \cite{6}  
\begin{equation}
\begin{aligned}
Z_{n,m}(f) = &-3 \sqrt{\Re\{Z_n\}\Re\{Z_{m}\}} \times \\
&\Bigg[ 
\frac{1}{2} \sin\alpha \sin\beta 
\left( \frac{1}{j k_0 d} + \frac{1}{(j k_0 d)^2} + \frac{1}{(j k_0 d)^3} \right) \\
&\quad + \cos\alpha \cos\beta 
\left( \frac{1}{(j k_0 d)^2} + \frac{1}{(j k_0 d)^3} \right) 
\Bigg] e^{-j k_0 d},
\end{aligned}
\label{eq:mutual}
\end{equation}
where $\Re\{Z_n\}$ and $\Re\{Z_{m}\}$ are the real parts of the self-impedance of the antennas obtained from (\ref{self}).

\begin{figure}
    \centering
    \includegraphics[width=0.5\linewidth]{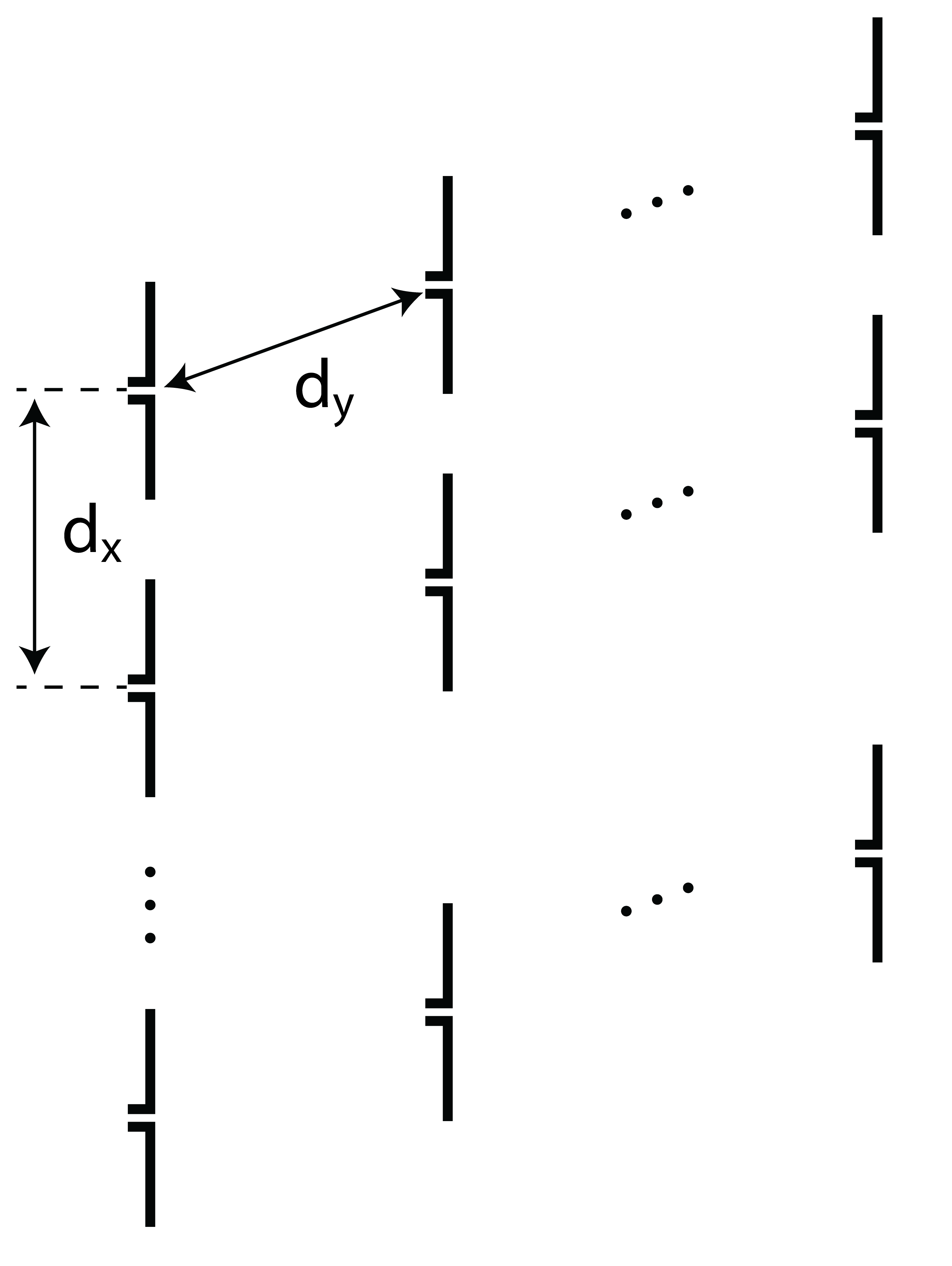}
    \caption{Rectangular array}
    \label{fig:4}
\end{figure}

Consider a rectangular planar array consisting of $N_x \times N_y$ antenna elements as shown in Fig.~\ref{fig:4},  
with inter-element spacings $d_x$ along the $x$-axis and $d_y$ along the $y$-axis.  
The array is assumed to lie in the $xy$-plane and centered at the origin,  
such that the position vector of the element at index $(n_x,n_y)$,  
where $n_x \in \{1,\dots,N_x\}$ and $n_y \in \{1,\dots,N_y\}$, is given by
\[
\mathbf{r}_{n_x,n_y} =
\begin{bmatrix}
\left(n_x-\tfrac{N_x+1}{2}\right)d_x \\
\left(n_y-\tfrac{N_y+1}{2}\right)d_y \\
0
\end{bmatrix}.
\]

The self-impedance of each element, corresponding to the diagonal entries of  
$\mathbf{Z}_{TT}$, can be directly obtained from \eqref{self}.  
For the mutual coupling between two distinct elements $(n_x,n_y)$ and $(n_x',n_y')$,  
we employ the general mutual-impedance expression \eqref{eq:mutual},  
where the inter-element distance is

\begin{equation}
\begin{split}
d_{(n_x,n_y),(n_x',n_y')} = 
\Big[ &\big((n_x'-n_x)d_x\big)^2 
      + \big((n_y'-n_y)d_y\big)^2 \\
      &+ z^2 \Big]^{1/2}.
\end{split}
\end{equation}
Here, $z$ represents the axial separation along the $z$-axis between the two rectangular arrays.  
For $\mathbf{Z}_{TT}$ and $\mathbf{Z}_{RR}$ (coupling within a same planar array), we set $z = 0$,
while for $\mathbf{Z}_{ST}$ (coupling between two parallel planar arrays),  
$z = d_z$, where $d_z$ denotes the distance along the $z$-axis between two arrays.

The orientation angles $\alpha_{(n_x,n_y),(n_x',n_y')}$ and $\beta_{(n_x,n_y),(n_x',n_y')}$ are computed  
from the relative geometry of the two elements along the line joining them and are given by
\[
\begin{aligned}
\cos\alpha_{(n_x,n_y),(n_x',n_y')} &= 
\frac{(n_x'-n_x)d_x}{d_{(n_x,n_y),(n_x',n_y')}} ,\\
\beta_{(n_x,n_y),(n_x',n_y')} &= 
\pi - \alpha_{(n_x,n_y),(n_x',n_y')} .
\end{aligned}
\]

By substituting these parameters into \eqref{eq:mutual},  
we obtain the off-diagonal entries of $\mathbf{Z}_{TT}$.  
Due to reciprocity, $\mathbf{Z}_{TT}$ is symmetric.  
Similarly, since the BS transmit array and the SIM array are located on two parallel planes  
separated by $d_z$ along the $z$-axis, the entries of $\mathbf{Z}_{ST}$  
can be obtained using the same formulation by setting $z = d_z$ in the distance computation.
\subsection{Model for SIM}
We consider a SIM implemented as an $L$-layer cascaded structure of transmissive RISs. Each transmissive RIS is modeled as two facing layers with a phase shifter embedded in between. In this configuration, the first layer receives the incident signal, the phase shift is applied, and the second layer subsequently re-transmits the modified signal to the next layer \cite{5}. We assume that the number of antenna elements in each layer is the same, so that each RIS consists of $2M$ ports. Consequently, the total number of ports in the SIM is $2ML$. The effect of SIM is described by $\mathbf{G} = \bold{(Z_{SS}+Z_{S})^{-1}}$  from Eq. (\ref{eq:transfer}), where $\mathbf{G}$ consists of ${2L \times 2L}$ sub-matrices $\bold{G}_{i,j} \in \mathbb{C}^{M \times M}$: 
\begin{equation}
\mathbf{G} =
\begin{bmatrix}
\mathbf{G}_{1,1} & \mathbf{G}_{1,2} & \cdots & \mathbf{G}_{1,2L} \\
\mathbf{G}_{2,1} & \mathbf{G}_{2,2} & \cdots & \mathbf{G}_{2,2L} \\
\vdots & \vdots & \ddots & \vdots \\
\mathbf{G}_{2L,1} & \mathbf{G}_{2L,2} & \cdots & \mathbf{G}_{2L,2L}
\end{bmatrix}.
\label{eq:G_matrix}
\end{equation}

In the case of $\mathbf{Z}_{SS}$ and $\mathbf{Z}_{S}$, both exhibit a banded structure and can be
decomposed into sub-matrices of size $M \times M$:
\begin{equation}
\mathbf{Z}_{SS}(f) =
\begin{bmatrix}
\mathbf{Z}_{2,2}^{(0)} & \mathbf{0}                 & \mathbf{0}                 & \mathbf{0}                 & \cdots & \mathbf{0} \\
\mathbf{0}              & \mathbf{Z}_{1,1}^{(1)}     & \mathbf{Z}_{1,2}^{(1)}     & \mathbf{0}                 & \cdots & \mathbf{0} \\
\mathbf{0}              & \mathbf{Z}_{2,1}^{(1)}     & \mathbf{Z}_{2,2}^{(1)}     & \mathbf{0}                 & \cdots & \mathbf{0} \\
\mathbf{0}              & \mathbf{0}                 & \mathbf{0}                 & \mathbf{Z}_{1,1}^{(2)}     & \cdots & \mathbf{0} \\
\vdots                  & \vdots                     & \vdots                     & \vdots                     & \ddots & \vdots \\
\mathbf{0}              & \mathbf{0}                 & \mathbf{0}                 & \mathbf{0}                 & \cdots & \mathbf{Z}_{1,1}^{(L)}
\end{bmatrix}.
\label{eq:ZSS_block}
\end{equation}

\begin{figure*}[t]
  \centering
  \includegraphics[width=0.9\textwidth]{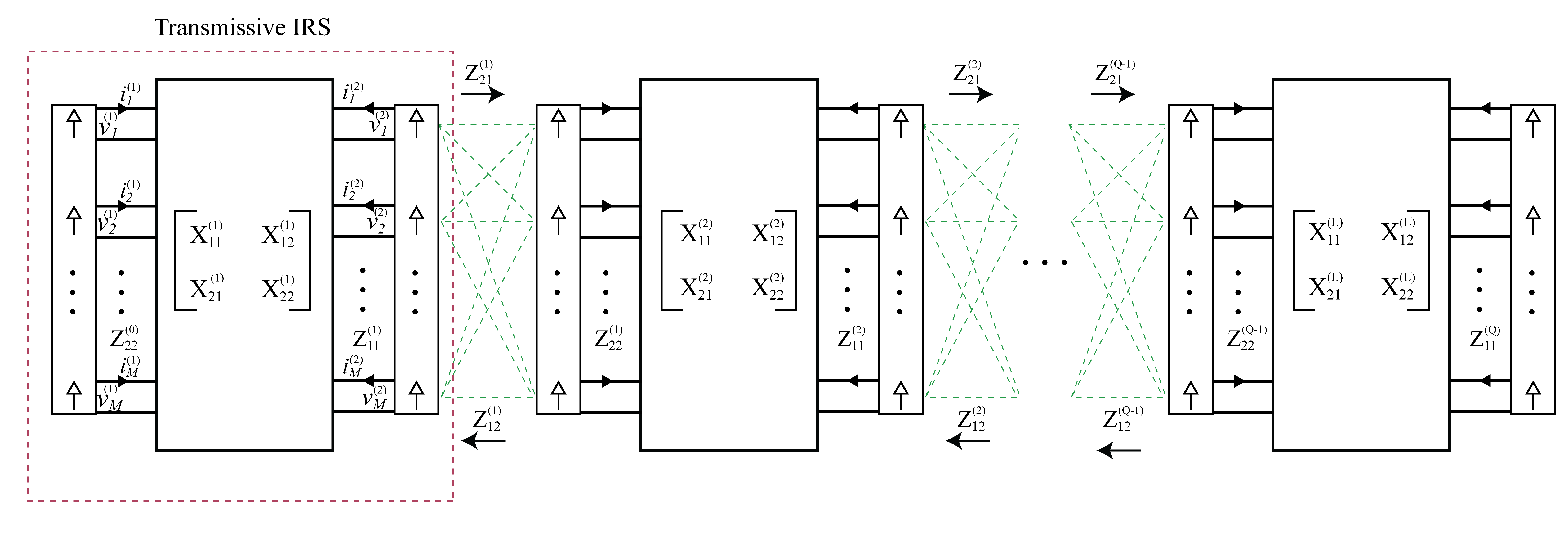}

  \caption{SIM model \cite{3}}
  \label{fig:5}
\end{figure*}

Inspired by \cite{3}, the SIM model is shown in Fig.~\ref{fig:5}. According to this model, the band matrix $\mathbf{Z}_{SS}$ can be decomposed into sub-matrices
$\mathbf{Z}^{(\ell)}_{i,j} \in \mathbb{C}^{M \times M}$, with $\ell = 1,2,\ldots,L-1$ and $i,j \in \{1,2\}$.
These $M \times M$ sub-matrices characterize the interconnections between consecutive layers of the SIM.
In particular, $\mathbf{Z}^{(\ell)}_{i,j}$ relates the $i$-th facing of the $2\ell$-th layer to the
$j$-th facing of the $(2\ell+1)$-th layer. We compute these blocks using the method introduced in Section~\ref{sec2:B}. Additionally, $\mathbf{Z}^{(0)}_{2,2} \in \mathbb{C}^{M \times M}$ characterizes the ports associated
with the first layer of the SIM, while $\mathbf{Z}^{(L)}_{1,1} \in \mathbb{C}^{M \times M}$ characterizes
the ports of the last layer.

The load matrix $\mathbf{Z}_{S}(\boldsymbol{\phi})$ can be decomposed into
sub-matrices $\mathbf{X}^{(\ell)}_{i,j}(\boldsymbol{\phi}_\ell) \in \mathbb{C}^{M \times M}$,
with $\ell = 1,2,\ldots,L$ and $i,j \in \{1,2\}$. These $M \times M$ blocks describe the load-network relations
within each SIM layer. In particular, $\mathbf{X}^{(\ell)}_{i,j}$ specifies the impedance block associated with
the $i$-th facing of the $(2\ell-1)$-th port set and the $j$-th facing of the $2\ell$-th port set of the same layer,
parameterized by the control vector $\boldsymbol{\phi}_\ell$. For simplicity, we exclude the explicit dependence on $\boldsymbol{\phi}_\ell$ in the notation. The global load matrix can be represented by block sub matrices given as
\begin{equation}
\mathbf{Z}_{S}(\boldsymbol{\phi}) =
\begin{bmatrix}
\mathbf{X}^{(1)}_{1,1} & \mathbf{X}^{(1)}_{1,2} & \mathbf{0} & \mathbf{0} & \cdots & \mathbf{0} & \mathbf{0} \\
\mathbf{X}^{(1)}_{2,1} & \mathbf{X}^{(1)}_{2,2} & \mathbf{0} & \mathbf{0} & \cdots & \mathbf{0} & \mathbf{0} \\
\mathbf{0} & \mathbf{0} & \mathbf{X}^{(2)}_{1,1} & \mathbf{X}^{(2)}_{1,2} & \cdots & \mathbf{0} & \mathbf{0} \\
\mathbf{0} & \mathbf{0} & \mathbf{X}^{(2)}_{2,1} & \mathbf{X}^{(2)}_{2,2} & \cdots & \mathbf{0} & \mathbf{0} \\
\vdots & \vdots & \vdots & \vdots & \ddots & \vdots & \vdots \\
\mathbf{0} & \mathbf{0} & \mathbf{0} & \mathbf{0} & \cdots & \mathbf{X}^{(L)}_{1,1} & \mathbf{X}^{(L)}_{1,2} \\
\mathbf{0} & \mathbf{0} & \mathbf{0} & \mathbf{0} & \cdots & \mathbf{X}^{(L)}_{2,1} & \mathbf{X}^{(L)}_{2,2}
\end{bmatrix}.
\label{eq:Zs_block}
\end{equation}

We assume that each layer of the SIM consists of diagonal phase shifts. Hence, for $i,j \in \{1,2\}$ and $\ell = 1,2,\ldots,L$, the sub-matrices $\mathbf{X}^{(\ell)}_{i,j}(\boldsymbol{\phi}_\ell)$ in \eqref{eq:Zs_block} are diagonal matrices whose entries correspond to the tunable load parameters. Specifically, we write
\begin{equation}
\mathbf{X}^{(\ell)}_{i,j}(\boldsymbol{\phi}_\ell) =
\begin{bmatrix}
x^{(\ell)}_{1}(i,j) & 0 & \cdots & 0 \\
0 & x^{(\ell)}_{2}(i,j) & \cdots & 0 \\
\vdots & \vdots & \ddots & \vdots \\
0 & 0 & \cdots & x^{(\ell)}_{M}(i,j)
\end{bmatrix},
\label{eq:X_diag}
\end{equation}
where $x^{(\ell)}_{m}(i,j)$ denotes the $m$-th diagonal entry, parameterized by 
the phase shift $\phi_{\ell,m}$ associated with the $m$-th meta-atom of the $\ell$-th layer.  {This diagonal structure corresponds to phase-only control, while any amplitude variations arise implicitly from inter-layer coupling and frequency-dependent effects captured by the circuit-theoretic formulation. By allowing $\mathbf{X}^{(\ell)}_{i,j}$ to be full matrices, the proposed framework can accommodate more general hardware models with joint phase–amplitude control using non-local interactions \cite{35}.} 

To characterize each diagonal entry $x^{(\ell)}_{m}(i,j)$, 
we model the $m$-th transmissive meta-atom as a two-port network with the following $S$-parameter matrix:
\begin{equation}
\mathbf{S}_{m}^{(\ell)} =
\begin{bmatrix}
0 & e^{j\phi_{\ell,m}} \\
e^{j\phi_{\ell,m}} & 0
\end{bmatrix},
\label{eq:S_matrix}
\end{equation}
which represents full transmission with tunable phase shift $\phi_{\ell,m}$.

Transforming \eqref{eq:S_matrix} into the Z-parameter domain, 
we obtain the corresponding two-port impedance matrix as follows:
\begin{equation}
\mathbf{Z}_{m}^{(\ell)} = j Z_0
\begin{bmatrix}
\dfrac{\cos(\phi_{\ell,m})}{\sin(\phi_{\ell,m})} & \dfrac{1}{\sin(\phi_{\ell,m})} \\
\dfrac{1}{\sin(\phi_{\ell,m})} & \dfrac{\cos(\phi_{\ell,m})}{\sin(\phi_{\ell,m})}
\end{bmatrix}.
\label{eq:Z_matrix}
\end{equation}

Consequently, each diagonal block $\mathbf{X}^{(\ell)}_{i,j}(\boldsymbol{\phi}_\ell)$ in \eqref{eq:X_diag} is obtained directly from the Z-parameter representation in \eqref{eq:Z_matrix}. 
In this formulation, the entire block is fully characterized by the set of phase shifts 
$\{\phi_{\ell,1}, \phi_{\ell,2}, \ldots, \phi_{\ell,M}\}$, which serve as the tunable load parameters 
of the $\ell$-th layer.

{Let us consider the case where inter-layer reflections and mutual coupling are ignored, and all ports are matched at reference impedance $Z_0$. In particular, let $\mathbf Z_{12}^{(l)}=\mathbf 0$ and $\mathbf Z_{11}^{(l)}=\mathbf Z_{22}^{(l-1)}=Z_0\mathbf I_M$ for all layer indices $l$. Under these assumptions the SIM model reduces to the conventional model:}

{\begin{equation}
\mathbf G_{2L,1}
=
\left(-\frac{1}{2Z_0}\right)^{L}
e^{j\boldsymbol\phi_L}
\prod_{l=L-1}^{1}\mathbf Z^{(l)}_{2,1} e^{j\boldsymbol\phi_l},
\end{equation}
where $\frac{1}{2Z_0}\mathbf Z^{(l)}_{2,1}$ is being modeled by the Rayleigh Sommerfeld diffraction equation in conventional models, which itself is questionable as it fails when antenna element size is not $\gg \lambda$.} 

Moreover, we assume there is no direct path between the transmitter and the receiver. Under these conditions, the transfer function in \eqref{eq:transfer} can be written as
$\bold{H_Z} = \bold{Z_{RS}(Z_{SS}+Z_{S})^{-1}Z_{ST}}$. Here,
$\bold{Z_{ST}}$ has only the first $M$ non-zero rows because only the first facing of the first layer is excited by the transmitter, while $\bold{Z_{RS}}$ has only the last $M$ non-zero columns since only the last facing layer of the SIM is connected to the receiver. Let $\bold{H}_1 \in \mathbb{C}^{M \times N_t}$ denote the matrix formed by the first $M$ rows of $\bold{Z_{ST}}$ and $\bold{H}_2 \in \mathbb{C}^{K \times M}$ denote the matrix formed by the last $M$ columns of $\bold{Z_{RS}}$. Then, the transfer function can be expressed as:
\begin{equation}
    \bold{H_Z} = \bold{H}_2 \bold{G}_{2L,1} \bold{H}_1.
\end{equation}

\subsection{Far Field Transimpedance Matrix}
{The external propagation environment from the last SIM layer to the users is incorporated through the far-field transimpedance matrix $\mathbf H_{2}(f)$, while $\mathbf H_{1}(f)$ characterizes the coupling between the BS transmit array and the first SIM layer.} As the users are in the far field, using circuit-theoretic principles together with the Friis transmission equation, the multipath (MP) transimpedance expression can be written as~\cite{6,34}:

\setlength{\multlinegap}{0pt} 
\begin{multline}
\hspace{-3pt}
\mathbf{H}^{\mathrm{}}_{2}(f)=
\frac{c\,\sqrt{G_S G_R}}{2\pi f\, d_{RS}^{\alpha/2}}\,
\operatorname{diag}\!\big(\Re\{\mathbf{Z}_{RR}(f)\}\big)^{1/2} \\
\times
\left(\sum_{e=1}^{E}
\mathbf{a}_R(\theta_{R,p})\,
\mathbf{a}_S^{\top}(\varphi_{S,p},\psi_{S,p})
\right)
\operatorname{diag}\!\big(\Re\{\mathbf{Z}_{11}^{L}(f)\}\big)^{1/2}\,
e^{-j\phi_0},
\label{eq:Zmp_RS}
\end{multline}
\vspace{-8pt}
\begin{equation*}
\phi_0
= \pi
- \arctan\!\left(\frac{2\pi f\, a_S}{c}\right)
- \arctan\!\left(\frac{2\pi f\, a_R}{c}\right).
\end{equation*}
Here, $\alpha$ is the path-loss exponent and $d_{RS,k}$ denotes the distance between the SIM and the $k$-th receiver, $k\in\{1,\ldots,K\}$. The SIM and the $k$-th receiver have gains $G_S$ and $G_{R,k}$, respectively. For the $e$-th propagation path toward user $k$, the angle of arrival at the (ULA) receiver is $\theta_{R,k,p}$, while $(\varphi_{S,k,p}, \psi_{S,k,p})$ denote the elevation and azimuth angles measured from the last layer of the $M_x \times M_y$ planar SIM antenna array. These angles are defined with respect to the broadside axes of the receive and SIM arrays, respectively. The diagonal factors $\operatorname{diag}(\Re\{\mathbf{Z}_{RR,k}(f)\})^{1/2}$ and $\operatorname{diag}(\Re\{\mathbf{Z}_{11}^{(L)}(f)\})^{1/2}$ embed the port-wise self-impedances at the $k$-th receiver and the last layer of SIM, and 
\begin{equation*}
\mathbf{a}_{R}(\theta_R)
= \big[\,1,\; e^{\,\mathrm{j}\tfrac{2\pi d_R}{\lambda}\cos\theta_R},\; \ldots,\;
e^{\,\mathrm{j}\tfrac{2\pi d_R}{\lambda}(N_R-1)\cos\theta_R}\big]^{\!\top}\;\,,
\end{equation*}
\begin{multline*}
\mathbf{a}_{S}(\varphi,\psi) = \\[2pt]
\begingroup\setlength{\arraycolsep}{2pt}
\left[
\begin{array}{c}
1\\
e^{\,j \tfrac{2\pi d_S}{\lambda}\sin\varphi\,\sin\psi}\\
\vdots\\
e^{\,j \tfrac{2\pi d_S}{\lambda}(M_x-1)\sin\varphi\,\sin\psi}
\end{array}
\right]
\otimes
\left[
\begin{array}{c}
1\\
e^{\,j \tfrac{2\pi d_S}{\lambda}\sin\varphi\,\cos\psi}\\
\vdots\\
e^{\,j \tfrac{2\pi d_S}{\lambda}(M_y-1)\sin\varphi\,\cos\psi}
\end{array}
\right],
\endgroup
\end{multline*}
where $d_S$ denotes the inter-element spacing of the SIM uniform planar array and $d_R$ denotes the inter-element spacing of the receive uniform linear array.
\subsection{Performance Evaluation Metrics}

We consider $\mathbf h_{i,k}\in\mathbb C^{1\times N_t}$, the $k$th row of $\mathbf{H_Z}$ as the effective downlink channel between the base station (BS) and user $k$ on subband $i$, and $\mathbf w_{i,k}\in\mathbb C^{1\times N_t}$ as the corresponding transmit beamforming vector. The BS employs linear precoding across all users and subbands to mitigate multiuser interference.

Inter-user interference is modeled as colored noise. Accordingly, the interference plus noise covariance matrix in the transmit domain for user $k$ on subband $i$ is given by
\begin{equation}
\mathbf C_{i,k}
=\sigma_i^{2}\mathbf I_{N_t}
+\sum_{\substack{j=1\\ j\neq k}}^{K}
\mathbf h_{i,j}^{H}\mathbf w_{i,j}\mathbf w_{i,j}^{H}\mathbf h_{i,j},
\qquad
\mathbf C_{i,k}\in\mathbb C^{N_t\times N_t}.
\label{eq:Cik_def}
\end{equation}
The achievable sum rate of the SIM-assisted system can then be written in determinant form as
\begin{align}
R
&=\sum_{i=1}^{N_f}\sum_{k=1}^{K} B_i
\bigg[
\log_{2}\!\Big|\mathbf C_{i,k}
+ \mathbf h_{i,k}^{H}\,
\mathbf w_{i,k}\mathbf w_{i,k}^{H}\,
\mathbf h_{i,k}\Big|
\nonumber\\[-2pt]
&\hspace{28mm}
-\log_{2}\!\big|\mathbf C_{i,k}\big|
\bigg].
\label{eq:R_det_form}
\end{align}

Since $\mathbf{C}_{i,k}$ is positive semi-definite, using eigendecomposition, it can be written as
$\mathbf C_{i,k}=\mathbf U_{i,k}\boldsymbol{\Lambda}_{i,k}\mathbf U_{i,k}^{H}$,
and we obtain the whitened channel using
$\widetilde{\mathbf h}_{i,k}
=\mathbf h_{i,k}\,\mathbf U_{i,k}\boldsymbol{\Lambda}_{i,k}^{-1/2}$.
Applying the matrix determinant lemma, the sum rate can be rewritten as
\begin{align}
R
&=\sum_{i=1}^{N_f}\sum_{k=1}^{K} B_i\,
\log_{2}\!\big|
 \mathbf{I}+ \widetilde{\mathbf h}_{i,k}^{H}\,\mathbf w_{i,k}\mathbf w_{i,k}^{H}\,\widetilde{\mathbf h}_{i,k}\big|.
\label{eq:R_scalar_form}
\end{align}

\section{Formulation of Optimization Problem  and Solution}

\subsection{Problem Formulation}
For each subband $i\in\{1,\ldots,N_f\}$, let 
$\mathbf{H}_{1,i}\in\mathbb{C}^{M\times N_t}$ denote the impedance matrix from the BS with $N_t$ antennas to the first facing of the SIM with $M=M_xM_y$ antenna elements per layer, and let 
$\mathbf{H}_{2,i}\in\mathbb{C}^{K\times M}$ denotes the impedance matrix from the last facing of the SIM to the $K$ single-antenna users.  
The $n$-th column of $\mathbf{H}_{1,i}$ is $\mathbf{h}_{1,i}^{(n)}\in\mathbb{C}^{M\times1}$, so
$\mathbf{H}_{1,i}=[\,\mathbf{h}_{1,i}^{(1)},\ldots,\mathbf{h}_{1,i}^{(N_t)}\,]$.  
The $k$-th row of $\mathbf{H}_{2,i}$ is $(\mathbf{h}_{2,i}^{(k)})^{\top}$ with $\mathbf{h}_{2,i}^{(k)}\in\mathbb{C}^{M\times1}$, hence
$\mathbf{H}_{2,i}=[\,(\mathbf{h}_{2,i}^{(1)})^{\top},\ldots,(\mathbf{h}_{2,i}^{(K)})^{\top}\,]^{\top}$.  
The SIM phase vector is $\boldsymbol{\phi}=[\phi_{1,1},\ldots,\phi_{L,M}]^{\top}$ for $L$ layers. Let $\mathbf{w}_{k,i}\in\mathbb{C}^{1\times N_t}$ be the transmit beamformer for user $k$ on subband $i$.  
The effective BS–user channel on subband $i$ is
\begin{equation}
\label{eq:hki_def_new}
\mathbf{h}_{k,i}
=\big(\mathbf{h}^{(k)}_{2,i}\big)^{\top}\,\mathbf{G}(\boldsymbol{\phi},f_i)\,\mathbf{H}_{1,i}
\in\mathbb{C}^{1\times N_t}.
\end{equation}

The constrained optimization problem to maximize the wideband sum-rate is given below:
\begin{equation}
\label{eq:obj_main_w}
\begin{aligned}
R^\star
= \max_{\mathbf{w}_{k,i},\,p_{k,i},\,\boldsymbol{\phi}}\quad
&\sum_{i=1}^{N_f}\sum_{k=1}^{K}
B_i\,\log_{2}\,\!\bigl|\mathbf{I}+\boldsymbol{\gamma}_{k,i}\bigr|,
\end{aligned}
\end{equation}
\addtocounter{equation}{-1}
\begin{subequations}\label{eq:constraints_w}
\begin{align}
\text{s.t.}\quad
& p_{k,i}= \operatorname{tr}\!\big(\mathbf{w}_{k,i}\mathbf{w}_{k,i}^{H}\big), \label{eq:cons_link_pw_w}\\
& \sum_{i=1}^{N_f}\sum_{k=1}^{K} p_{k,i}\le P_{\mathrm{tot}},\qquad p_{k,i}\ge 0, \label{eq:cons_power_budget}\\
& 0\le \phi_{\ell,m}<2\pi,  \label{eq:cons_phi}
\end{align}
\end{subequations}
\quad \quad \quad \quad \quad where
\begin{equation}
\label{eq:sinr_w_scalar}
\boldsymbol{\gamma}_{k,i}
= \widetilde{\mathbf h}_{i,k}^{H}\,\mathbf w_{i,k}\mathbf w_{i,k}^{H}\,\widetilde{\mathbf h}_{i,k}.
\end{equation}

\subsection{Proposed Solution of the Optimization Problem}
The problem defined in the previous section is non-convex. To obtain a solution for the problem, we adopt an Alternating Optimization (AO) algorithm that decomposes the problem into sub-problems, as described below: 

\begin{equation}
\label{prob:P1}
{
\begin{aligned}
\textbf{P1:}\;\; \operatorname*{\arg\max}_{\{p_{i,k},\mathbf w_{k,i}\}} \quad
& \sum_{i=1}^{N_f}\sum_{k=1}^{K}
B_i \log_2\,\!\Bigl|\mathbf{I}+\, \widetilde{\mathbf h}_{i,k}^{H}\,\mathbf w_{i,k}\mathbf w_{i,k}^{H}\,\widetilde{\mathbf h}_{i,k}\Bigr| \\
\text{s.t.}\quad
& p_{k,i}= \operatorname{tr}\!\big(\mathbf w_{k,i}\mathbf w_{k,i}^{H}\big),\ \forall\,i,k,\\
& \sum_{i,k} p_{k,i} \le P_{\mathrm{tot}}, \qquad p_{k,i}\ge 0,
\end{aligned}}
\end{equation}

\vspace{-12pt}
\begin{equation}
\label{prob:P2}
{
\begin{aligned}
\textbf{P2:}\;\; \operatorname*{\arg\max}_{\boldsymbol{\phi}} \quad
& \sum_{i=1}^{N_f}\sum_{k=1}^{K}
B_i \log_2\,\!\Bigl|\mathbf{I}+\, \widetilde{\mathbf h}_{i,k}^{H}\,\mathbf w_{i,k}\mathbf w_{i,k}^{H}\,\widetilde{\mathbf h}_{i,k}\Bigr| \\
\text{s.t.}\quad
& 0 \le \phi_{\ell,m} < 2\pi .
\end{aligned}}
\end{equation}


In this AO approach, the beamforming vectors and the phase shifts are optimized alternately. This is a widely adopted strategy since it effectively decouples a non-convex problem into subproblems that can be solved iteratively with reduced computational overhead. 


\subsubsection{Power Allocation and Beamforming Optimization}

The subproblem \textbf{P1} maximizes the wideband sum rate with fixed SIM phases $\boldsymbol{\phi}$. We use the iterative water–filling approach of \cite{7} and adapt it to the multi-band setting in \cite{8}.

For user $k$ on subband $i$, all other users are treated as interference $\mathbf C_{i,k}$ given in \eqref{eq:Cik_def}. 
The corresponding scalar gain is
$\lambda_{k,i}
=\|\widetilde{\mathbf h}_{k,i}\|^{2}
=\mathbf h_{k,i}\mathbf C_{i,k}^{-1}\mathbf h_{k,i}^{H}$.
Let $\mathbf Q_{k,i}\,\boldsymbol{\Delta}_{k,i}^{1/2}\,\mathbf R_{k,i}^{H}$ and $\mathbf S_{k,i}\,\boldsymbol{\nabla}_{k,i}^{1/2}\,\mathbf T_{k,i}^{H}$ denote the singular value decompositions of $\widetilde{\mathbf h}_{k,i}^{H}$ and $\mathbf w_{k,i}^{H}$, respectively, where $\mathbf Q_{k,i},\mathbf R_{k,i},\mathbf S_{k,i},\mathbf T_{k,i}$ are unitary matrices. We maximize the objective function by choosing $\mathbf S_{k,i}=\mathbf Q_{k,i}$. Hence, the maximization reduces to
\begin{equation}
\label{eq:WF_gain_form_final}
{
\begin{aligned}
\operatorname*{\arg\max}_{\{p_{k,i}\}} \;&
\sum_{i=1}^{N_f}\sum_{k=1}^{K} B_i\,\log_{2}\!\bigl(1+\lambda_{k,i}\,p_{k,i}\bigr)\\
\text{s.t.}\;&
\sum_{i=1}^{N_f}\sum_{k=1}^{K} p_{k,i}\le P_{\mathrm{tot}},\quad p_{k,i}\ge 0.
\end{aligned}}
\end{equation}

\noindent
We initialize with any feasible $\{p_{k,i}^{(0)}\}$. At iteration $t$, for each user–sub-band pair we construct $\mathbf C_{i,k}^{(t)}$ from the current powers, compute $\widetilde{\mathbf h}_{k,i}^{(t)}$, and set $\lambda_{k,i}^{(t)}=\|\widetilde{\mathbf h}_{k,i}^{(t)}\|^{2}$. We update the power by iterative water–filling given as:
\begin{equation}
    p_{k,i}^{(t+1)}=[\mu^{(t)}-1/\lambda_{k,i}^{(t)}]^+,
\end{equation}
where $\mu^{(t)}$ enforces $\sum_{i=1}^{N_f}\sum_{k=1}^{K} p_{k,i}^{(t+1)}=P_{\mathrm{tot}}$. The operator $[x]^+$ returns $x$ if $x\ge 0$ and $0$ otherwise.
After the power update, the optimum beamforming can be obtained by:
\[
\mathbf (w_{k,i}^{(t+1)})^T
= \mathbf Q_{k,i}
\begin{bmatrix}
\sqrt{p_{k,i}^{(t+1)}}\\[2pt]
\mathbf 0
\end{bmatrix}.
\]

This optimization process inherently performs carrier aggregation over the wideband spectrum. Through the water-filling mechanism, transmit power is selectively distributed across subbands according to their instantaneous channel gains and interference conditions. Consequently, only subbands that contribute meaningfully to the overall spectral efficiency are allocated nonzero power, while those with unfavorable propagation conditions remain inactive. This adaptive subband activation effectively emulates the carrier aggregation behavior observed in practical multiband systems, thereby ensuring efficient utilization of spectral resources within the proposed SIM-assisted wideband transmission system.

\subsubsection{Phase Shift Optimization}
With the fixed beamforming matrix, we maximize the sum-rate with respect to the SIM phases
$\boldsymbol{\phi}$:
{
\begin{multline}
\mathcal{J}(\boldsymbol{\phi})
=\sum_{i=1}^{N_f}\sum_{k=1}^{K}
B_i\,\log_2\!\bigl(1+s_{i,k}(\boldsymbol{\phi})\bigr), \mbox{where}\\
s_{i,k}(\boldsymbol{\phi})
= p_{k,i}\,\mathbf{h}_{k,i}\,\mathbf{C}_{i,k}^{-1}\,\mathbf{h}_{k,i}^{H}.
\label{eq:P2_obj_phi}
\end{multline}}
The effective channel is given as
{
\begin{align}
\mathbf{h}_{k,i}
&=(\mathbf{h}^{(k)}_{2,i})^{\top}\,\mathbf{G}(\boldsymbol{\phi},f_i)\,\mathbf{H}_{1,i}.
\label{eq:def_hki_phi}
\end{align}}
To optimize our phase shift, we use gradient ascent with step size $\eta_t>0$:
\begin{equation}
\phi_{\ell,m}^{(t+1)}
=\phi_{\ell,m}^{(t)}
+\eta_t\,
\Bigl[\frac{\partial \mathcal{J}}{\partial \phi_{\ell,m}}\Big],
\label{eq:phi_update}
\end{equation}
where $\nabla_{\boldsymbol{\phi}}\mathcal{J}(\boldsymbol{\phi})$ is the gradient of the objective function. 
To obtain the gradient, we first define the SIM-induced block on subband $i$ as
$\mathbf{H}_i(\boldsymbol{\phi})=(\mathbf{H}_{2,i})^{\top}\mathbf{G}(\boldsymbol{\phi},f_i)\mathbf{H}_{1,i}$.
Let $\mathbf{F}_{\ell,m}^{i}(\boldsymbol{\phi})$ denote the derivative of $l$-th layer and $m$-th element in $\mathbf G$ and the overall derivative is defined as
\begin{align}
\boldsymbol{\Delta}_{i,k}
&= -\Bigl[(\mathbf H_{2,i})^{\top}\,\mathbf F_{\ell,m}^{i}(\boldsymbol{\phi})\,\mathbf H_{1,i}\Bigr]_{k,:},
\label{eq:def_Delta}
\end{align}
\vspace{-5pt}
Let
\begin{equation}
\mathbf v_{i,k}=\mathbf C_{i,k}^{-1}\mathbf h_{k,i}^{H},
\qquad
\mathbf E_{i,k}=\mathbf v_{i,k}\mathbf v_{i,k}^{H}.
\label{eq:def_v_E}
\end{equation}
The derivative of the
SINR $s_{i,k}$ with respect to $\phi_{\ell,m}$ is given by (see \textbf{Appendix}):
{
\begin{multline}
\frac{\partial s_{i,k}}{\partial \phi_{\ell,m}}
= p_{i,k}\Biggl[
2\,\Re\!\bigl\{\boldsymbol{\Delta}_{i,k}\,\mathbf v_{i,k}\bigr\}
\\[-2pt]
\hspace{32pt}
-\,2\sum_{k'\neq k} p_{i,k'}\,
\Re\!\bigl\{\bigl(\mathbf E_{i,k}\mathbf h_{k',i}^{H}\bigr)^{H}\,\boldsymbol{\Delta}_{i,k'}\bigr\}
\Biggr].
\label{eq:dsik_final}
\end{multline}
}
Combining this with the gradient of  \eqref{eq:P2_obj_phi} gives the gradient of the objective function:
\begin{equation}
\label{eq:gradJ_final_phi}
\frac{\partial \mathcal{J}}{\partial \phi_{\ell,m}}
=
\sum_{i=1}^{N_f}\sum_{k=1}^{K}
\frac{B_i}{\ln 2}\,
\frac{1}{1+s_{k,i}}\,
\frac{\partial s_{k,i}}{\partial \phi_{\ell,m}}.
\end{equation}
We update the phase shifts using \eqref{eq:phi_update} to maximize the data rate.

\subsubsection{Convergence and Complexity}

The proposed AO algorithm exhibits a monotonic behavior, where each update of the beamforming and phase shift variables ensures a non-decreasing objective function value. The iterations proceed until the change in the objective value between two consecutive iterations becomes sufficiently small. Specifically, the convergence criterion is defined as \(|\mathcal{J}^{(t+1)} - \mathcal{J}^{(t)}| / |\mathcal{J}^{(t)}| < \epsilon\), where \(\epsilon\) is a small positive threshold (\(10^{-7}\)). 

During the beamforming optimization stage, the steps corresponding to $B_1$ through $B_8$ in Table~\ref{Table:1} are computed only once, as the equivalent channel is assumed to remain fixed in this stage. Subsequently, for each of the \(I_{P1}\) beamforming iterations, the updates in \(B_9\) and \(B_{10}\) are executed. Once beamforming is optimized, it is kept constant, and the phase shift is updated. In this stage, the equivalent channel \(\widetilde{\mathbf{h}}_{i,k}\) is recomputed at each iteration, and the operations corresponding to \(B_{11}\), \(B_{12}\), and \(B_{13}\) are used. \(I_{P2}\) iterations are used to perform this stage. Therefore, the total computational complexity of the proposed alternating optimization algorithm can be expressed as \((B_1 + B_2 + B_3 + B_4 + B_5 + B_6 + B_7 + B_8  + B_{11} + B_{12} + B_{13})I_{P2}+ (B_9 + B_{10})I_{P1},\) which represents the overall number of mathematical operations required to complete one full iteration of the alternating optimization procedure.

\begin{table*}[t]
\centering
\renewcommand{\arraystretch}{1.2} 
\caption{Number of mathematical operations required for different steps in optimization}
\begin{tabular}{|c|c|}
\hline
\textbf{Step} & \textbf{Number of Mathematical Operations}
\label{Table:1}
\\
\hline
$\mathbf{G}_{:,1}$ and $\mathbf{G}_{2L,:}$ 
&
$\displaystyle (18\,L - 6)\,N_f\,M^3 + (2\,N_f - 1)\,N\,K$
\hfill\;\textbf{\}}$B_1$
\\
\hline
$\mathbf{H}_Z$ 
&
$\displaystyle 2\,N_t\,(M^2 + K\,M)\,N_f$
\hfill\;\textbf{\}}$B_2$
\\
\hline
$\mathbf{F}$ 
&
$\displaystyle (3\,M^2 + 6\,M)\,L\,M\,N_f$
\hfill\;\textbf{\}}$B_3$
\\
\hline
$\dfrac{\partial \mathbf{H}_Z}{\partial \boldsymbol{\phi}}$ 
&
$\displaystyle 2\,N_t\,(M^2 + K\,M)\,L\,M\,N_f$
\hfill\;\textbf{\}}$B_4$
\\
\hline
$\mathbf{C}_{i,k}$
&
$\displaystyle
\big(\big((2N_t - 1) + (2N_t^2 + 1)\big)(K - 1) + N_t^2\big)\,K\,N_f$
\hfill\;\textbf{\}}$B_5$
\\
\hline
$\mathbf{U}_{i,k}\text{ and }\boldsymbol{\Lambda}_{i,k}$ 
&
$\displaystyle \mathcal{O}(N_t^4)\,K\,N_f$
\hfill\;\textbf{\}}$B_6$
\\
\hline
$\widetilde{\mathbf{h}}_{i,k}$
&
$\displaystyle K\,(2N_t^{2} + N_t)\,N_f$
\hfill\;\textbf{\}}$B_7$
\\
\hline
$s_{i,k}$
&
$\displaystyle 2\,K\,N_t\,N_f$
\hfill\;\textbf{\}}$B_8$
\\
\hline
Power allocation
&
$\displaystyle 3\,N_f\,K - 1$
\hfill\;\textbf{\}}$B_9$
\\
\hline
$\mathbf{w}_{k,i}^{(t+1)}$ update
&
$\displaystyle 2\,K\,N_t\,N_f$
\hfill\;\textbf{\}}$B_{10}$
\\
\hline
$\dfrac{\partial s_{k,i}}{\partial \phi_{\ell,m}}$ 
&
$\displaystyle N_f\,K\,\big(2N_t + (K - 1)(2N_t^2 + 2N_t)\big)$
\hfill\;\textbf{\}}$B_{11}$
\\
\hline
$\nabla_{\boldsymbol{\phi}}\mathcal{J}(\boldsymbol{\phi})$
&
$\displaystyle 2\,L\,M\,K\,N_f$
\hfill\;\textbf{\}}$B_{12}$
\\
\hline
Phase update $\boldsymbol{\phi}^{(t+1)}$
&
$\displaystyle 2\,L\,M$
\hfill\;\textbf{\}}$B_{13}$
\\
\hline
\end{tabular}
\end{table*}

\subsubsection{Partially Reconfigurable SIM}
As the number of antenna elements increases, the overall system complexity and control overhead of fully reconfigurable SIM and RIS architectures grow significantly. {Each additional element increases the number of control bits that must be transmitted whenever the metasurface configuration is updated}, which makes real-time optimization challenging in large-scale deployments. To address this issue, we consider a partially reconfigurable SIM design that leverages the flexibility of SIM while reducing the complexity of reconfiguration. In this configuration, the first layer is optimized once for a given channel realization and then kept fixed, { so that it provides a stable wavefront transformation, while only the last layer is adaptively reconfigured to compensate for residual multi-user interference and wideband phase variations}. This approach substantially reduces the number of control parameters and the corresponding signaling updates, thereby improving the system’s operational efficiency.

To quantify the impact of control overhead, we evaluate system performance in terms of goodput, which accounts for both the achievable rate and the signaling cost. The goodput is expressed as \cite{27}:
\begin{equation}
    G = \left(1 - \frac{b M_{\text{total}}}{\eta N_s}\right) R,
\end{equation}
where the penalty term reflects the fraction of each slot used for metasurface control signaling. Here, $M_{\text{total}}$ denotes the number of reconfigurable elements that are updated in a slot and each antenna element is controlled by $b$ number of bits. Therefore, updating $M_{\text{total}}$ elements requires $bM_{\text{total}}$ control bits per slot. Moreover, $\eta$ denotes the spectral efficiency of the control channel and $N_s$ denotes the number of transmitted symbols per slot, {so that $\frac{b M_{\text{total}}}{\eta N_s}$ represents the normalized signaling cost}. For a fully reconfigurable SIM, $M_{\text{total}} = LM$, where $L$ and $M$ denote the number of layers and elements per layer, respectively. In contrast, for the partially reconfigurable configuration where only the last layer is adjustable, $M_{\text{total}} = M$, which reduces the required control signaling proportionally by a factor of $L$.

{In addition to the reduction in control bits, the computational complexity of the optimization is also reduced. In particular, at each iteration, the number of mathematical operations required to compute $B_3$, $B_4$, $B_{12}$, and $B_{13}$ in Table~\ref{Table:1} is reduced by a factor of $L$.}


\section{Simulation Results and Analysis}

\subsection{Simulation Parameters}
{In this section, Monte Carlo simulation results are presented to numerically evaluate the developed model.} We consider a wideband system of total bandwidth $15\,\mathrm{GHz}$ centered at $f_c = 27\,\mathrm{GHz}$, and {$R$ denotes the system wide achievable sum rate (sum spectral efficiency) in $\mathrm{bps/Hz}$.} The band is partitioned into $N_f$ subbands of equal width. The total transmit power at the BS is $P_T = 50\mathrm{W}$, and the antenna element radius is $a = 0.0027\mathrm{m}$ ($\lambda_c/4$). The receiver noise power is set to $\sigma^2 = -70\mathrm{dB}$ unless mentioned otherwise. The distance between the BS transmit array and the last layer of the SIM is set to {$5\lambda_c$, where $\lambda_c$ denotes the wavelength at the center frequency.}
The distance between two adjacent antennas is set to $d_x = d_y = 5.56 \times 10^{-3}\mathrm{m}$, {which corresponds to half the wavelength at the center frequency, i.e., $d_x = d_y = \lambda_c/2$. Since the system operates over a wide bandwidth, the electrical spacing becomes frequency dependent across subbands, which naturally captures wideband effects such as beam squint and frequency-dependent array responses. Single-antenna users are randomly placed in the 3D region in front of the SIM, with their distances from the last SIM layer uniformly distributed between $10$~meters and $20$~meters.} The number of BS transmit antennas is set equal to the number of users. The path loss exponent is $\alpha = 2$, and the number of multipaths is $E = 1$. We set $b = 2$, the spectral efficiency of the control channel to $\eta = 2$, and $N_s = 700$.

\subsection{Simulation Results}

\begin{figure}
    \centering
    \includegraphics[width=0.98\linewidth]{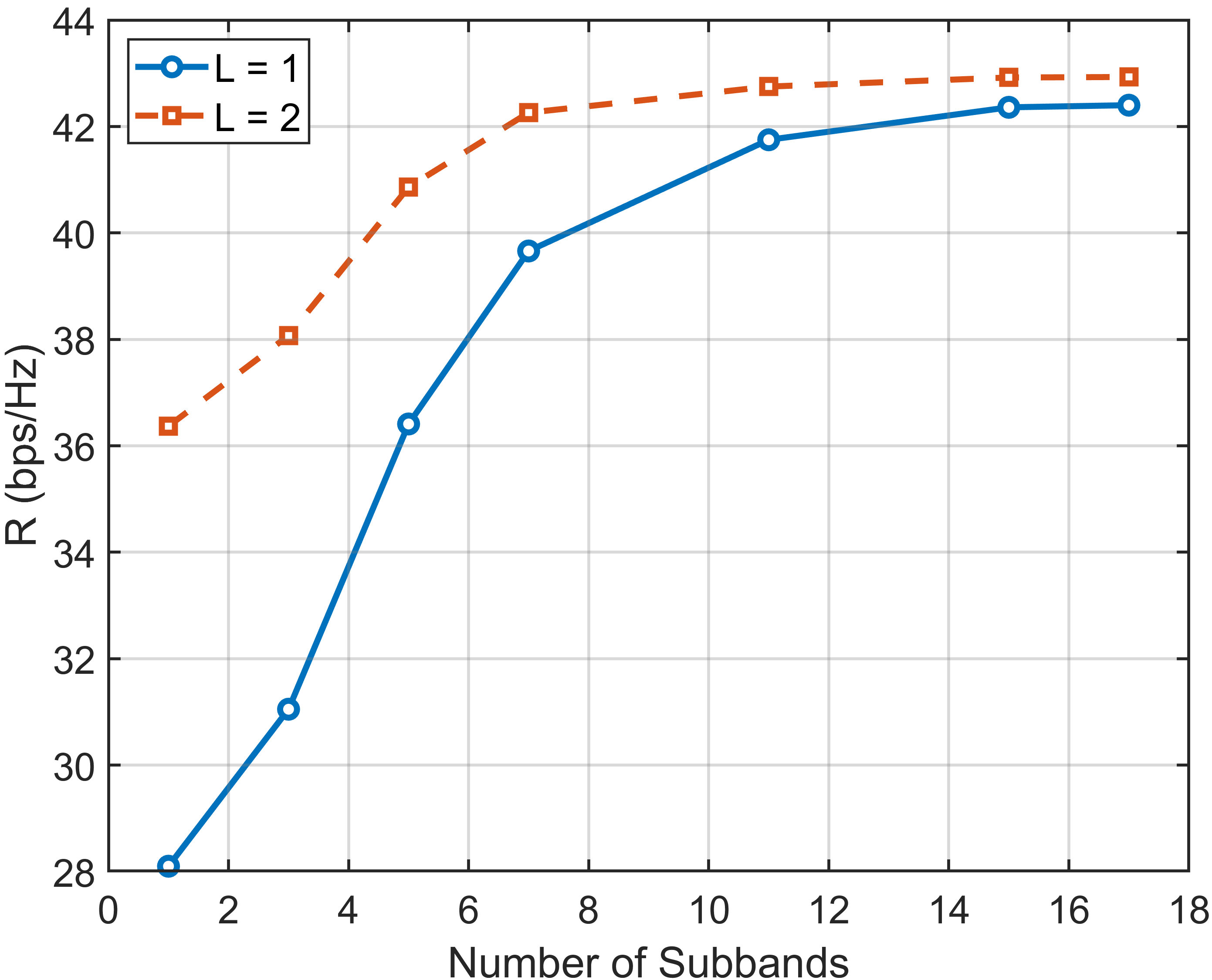}
    \caption{Variation of optimal spectral efficiency with the number of subbands for $K = 5$}
    \label{fig:6}
\end{figure}

We first investigate the effect of optimizing the phase shifts across multiple subbands over a wideband channel under uniform power allocation, on the performance of the RIS. Fig. ~\ref{fig:6} illustrates the influence of the number of subbands on the spectral efficiency for a system employing a total of 100 antenna elements. The single-layer RIS configuration is designed as a $10 \times 10$ array. For the multi-layer case ($L = 2$), each layer consists of a $10 \times 5$ metasurface, ensuring that the total number of antenna elements remains constant across both configurations.

The results show that increasing the number of subbands significantly enhances the spectral efficiency, as the optimization captures more accurate frequency selective channel behavior. At lower subband counts, the multi-layer metasurface ($L = 2$) exhibits superior performance compared to the single-layer configuration due to its cascaded structure. However, as the number of subbands increases, the performance gap between the single-layer and multi-layer configurations narrows. Beyond approximately $15$ subbands, the achievable spectral efficiency tends to saturate, indicating that the wideband optimization has effectively converged and that further subband divisions yield negligible additional gain.

\begin{figure}[t]
\centering

\subfloat[\label{fig:7}]{%
\includegraphics[width=0.48\linewidth]{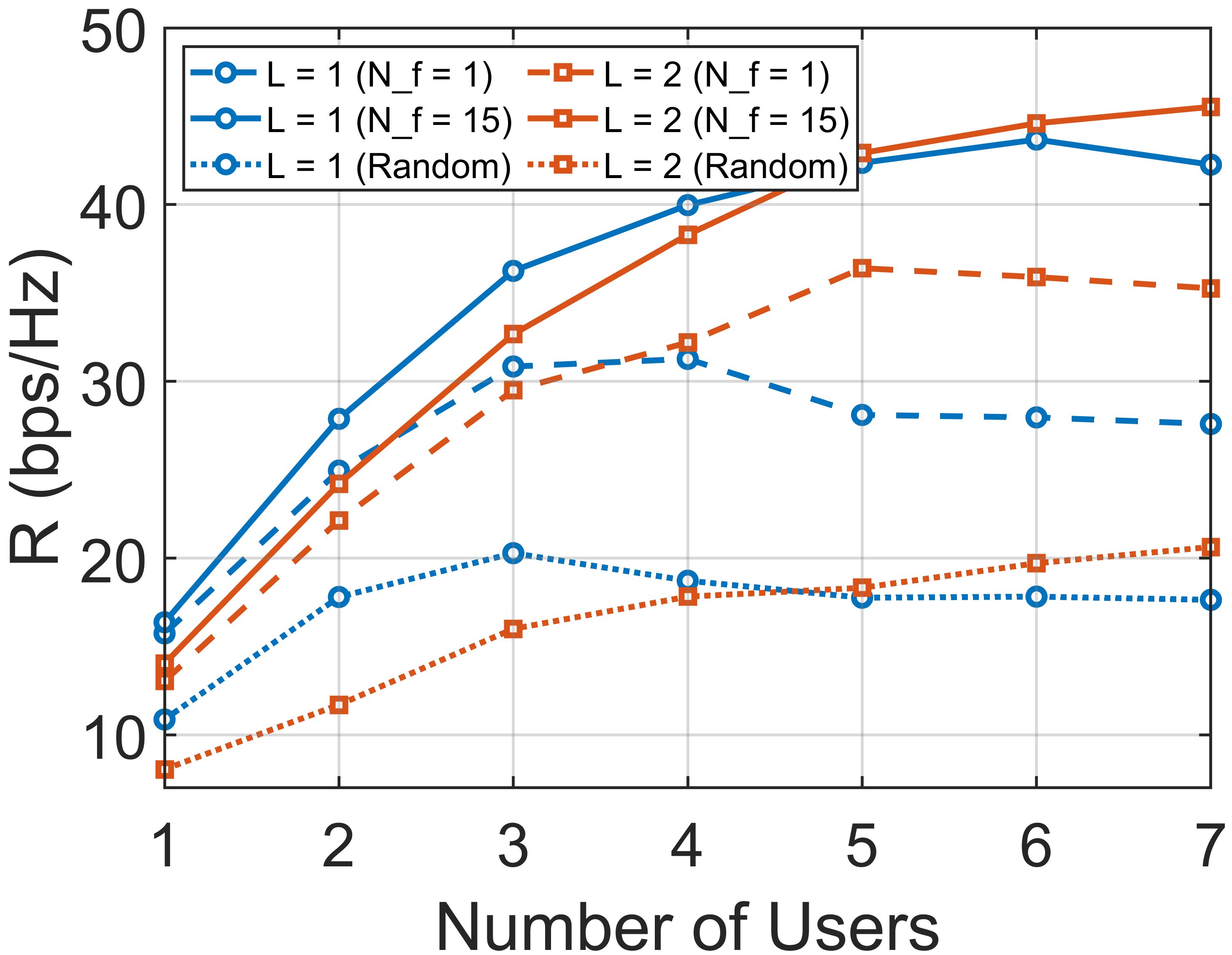}
}
\hfill
\subfloat[\label{fig:8}]{%
\includegraphics[width=0.48\linewidth]{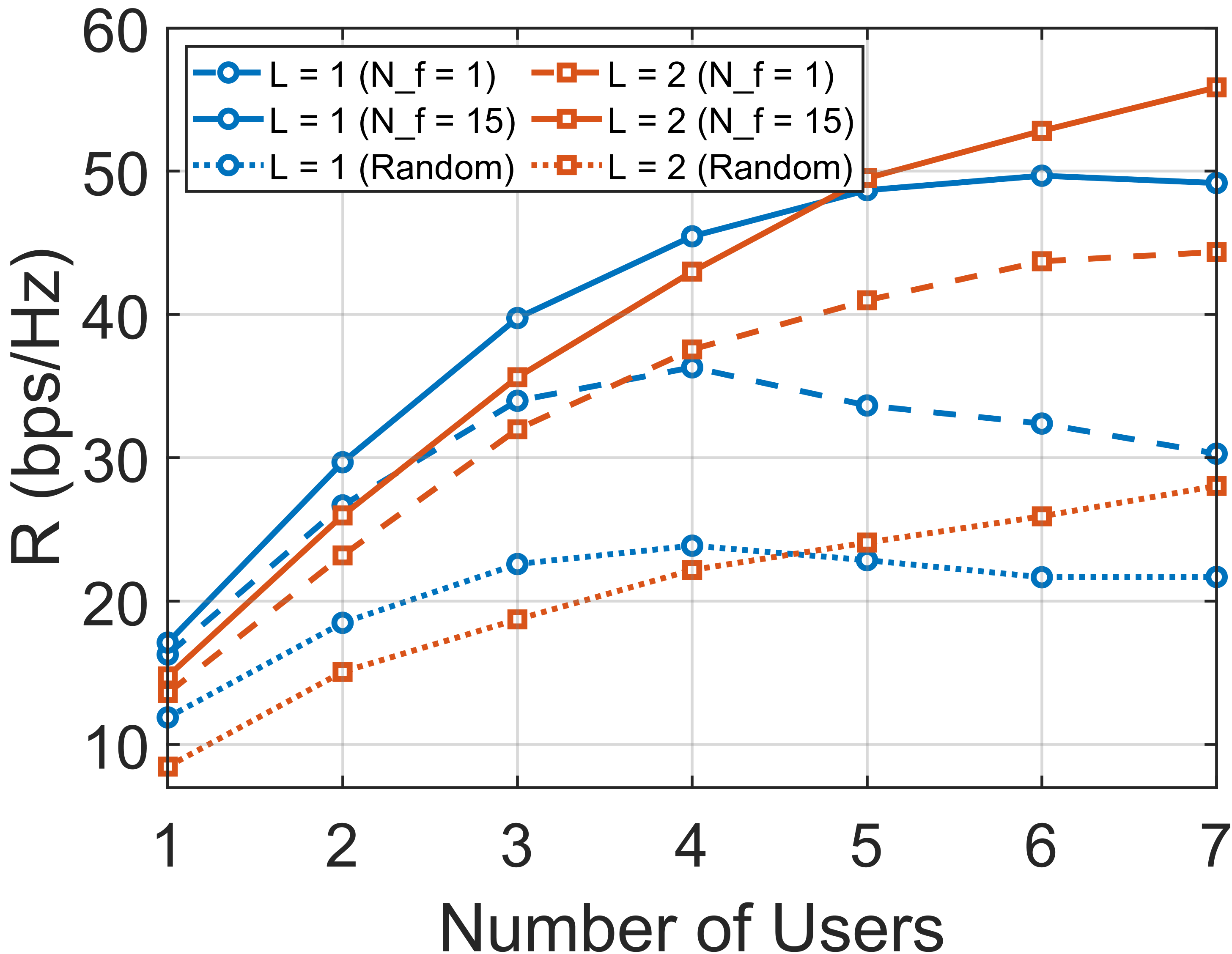}
}

\vspace{1ex}

\subfloat[\label{fig:9}]{%
\includegraphics[width=0.48\linewidth]{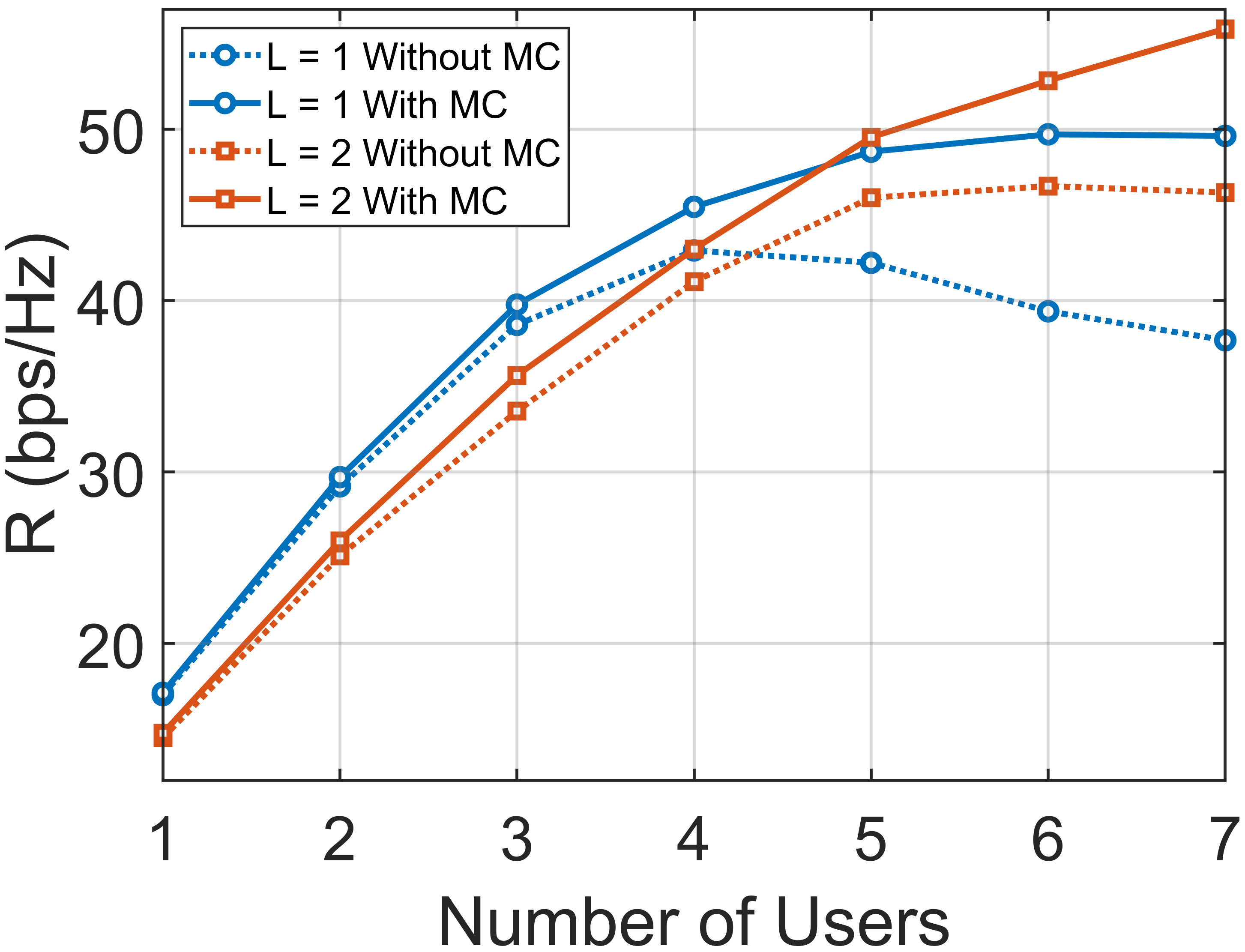}
}
\hfill
\subfloat[\label{fig:11}]{%
\includegraphics[width=0.48\linewidth]{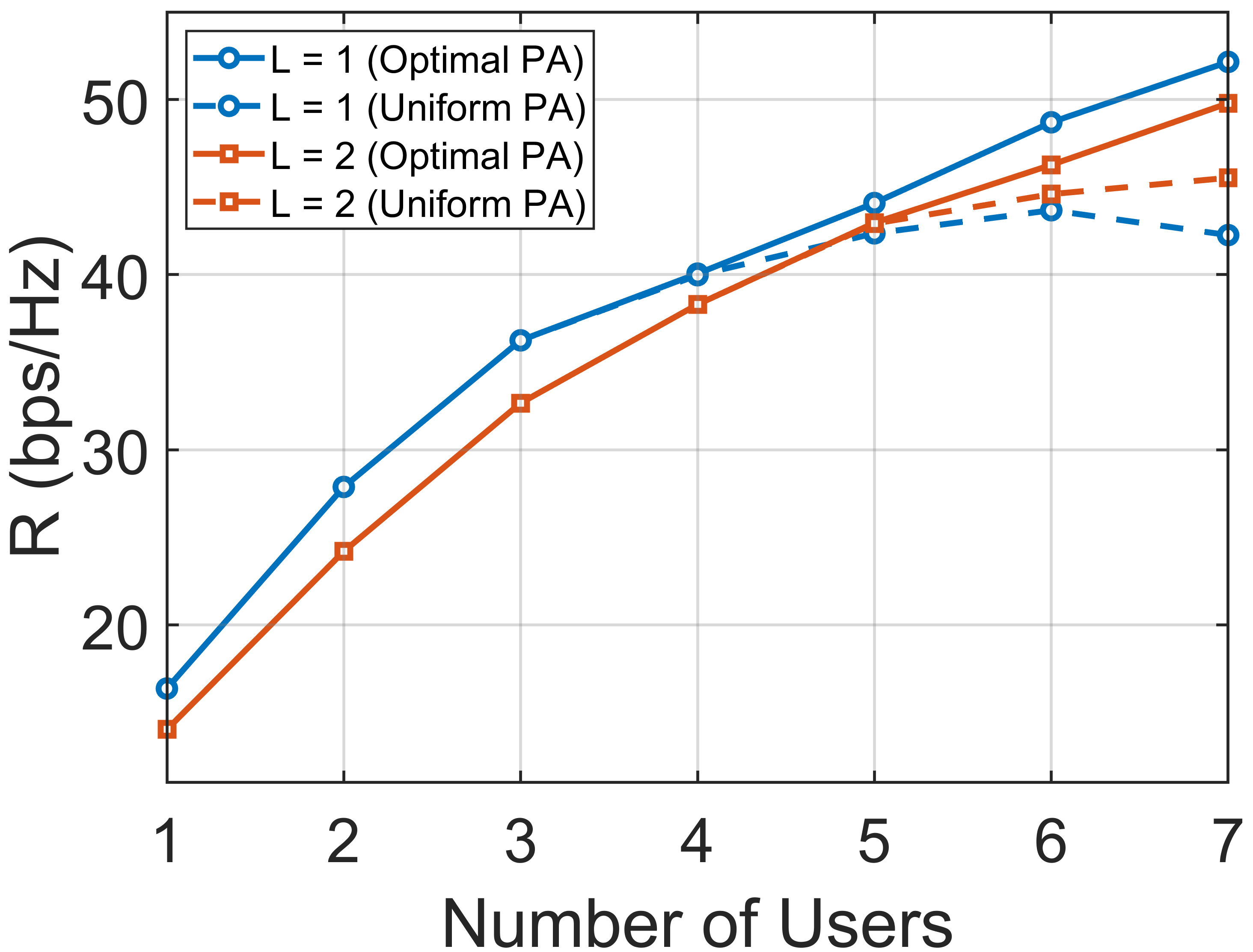}
}

\caption{{Spectral efficiency versus the number of users for different metasurface configurations:
(a) $M_{\mathrm{total}} = 100$,
(b) $M_{\mathrm{total}} = 144$,
(c) with and without mutual coupling,
(d) with and without power allocation}}
\label{fig:merged}
\end{figure}

Next, we investigate the effect of the number of users on the performance of the single-layer RIS and the multi-layer structure (SIM). Fig.~\ref{fig:7} illustrates the variation in spectral efficiency with the number of users for a total of $100$ antenna elements. For a small number of users, where inter-user interference is relatively low, the single-layer RIS exhibits better performance due to its larger effective aperture area. However, as the number of users increases, inter-user interference becomes more dominant. In this regime, the single-layer structure struggles to generate sufficiently complex beam patterns capable of effectively separating users. In contrast, the multi-layer SIM, due to its additional spatial degrees of freedom, begins to outperform the single-layer RIS while maintaining the same total number of elements. It is also observed that beyond approximately $K=5$ users, the SIM achieves higher spectral efficiency than the RIS under wideband operation with the same total number of antenna elements. Moreover, the performance gap between single-subband ($N_f=1$) and multiple-subband ($N_f=15$) optimization widens as the number of users increases, highlighting the increasing frequency selectivity of the channel in multi-user wideband scenarios. This demonstrates the necessity of optimizing the metasurface phase shifts across multiple subbands to fully exploit the wideband potential of metasurface-assisted systems.

Similarly, Fig.~\ref{fig:8} presents the results for a larger RIS consisting of $144$ antenna elements. As the size of the RIS increases, the difference in spectral efficiency between the single-layer and multi-layer configurations becomes more pronounced, especially at higher user counts. This behavior arises because larger RISs exhibit stronger frequency-selective responses. Therefore, employing a cascaded (multi-layer) structure proves more beneficial than merely increasing the number of elements within a single layer, as the additional depth provides a more effective way to enhance spectral efficiency.

In Fig.~\ref{fig:9}, for a total of $144$ antenna elements, we examine the impact of incorporating mutual coupling and mutual reflections in the RIS model. For a small number of users, the effect of ignoring mutual coupling is relatively insignificant, as inter-user interference and element interactions remain limited. However, as the number of users increases, the impact of mutual coupling becomes more pronounced, particularly in the single-layer configuration. The single-layer RIS, having a larger effective aperture area, is more susceptible to distortions introduced by coupling among closely spaced elements. When these effects are properly modeled and taken into account, the system achieves noticeably higher spectral efficiency. This highlights the importance of accurately accounting for mutual coupling in wideband RIS-assisted systems, especially under higher user densities or closely spaced antenna elements.

In Fig.~\ref{fig:11}, we illustrate the impact of employing optimal power allocation (PA) at the base station on the spectral efficiency of both the single-layer RIS and the multi-layer SIM. For the RIS, the effect of power allocation becomes significant beyond $K=4$ users, where the combination of active beamforming at the transmitter and passive beamforming at the metasurface enables it to outperform the SIM with the same total number of antenna elements. For the SIM, power allocation provides noticeable improvement after approximately $K=5$ users, as the system benefits from additional degrees of freedom for reducing interference. Nevertheless, when both active and passive beamforming are jointly optimized, the single-layer RIS consistently achieves higher spectral efficiency than the multi-layer SIM under the same total number of antenna elements. This is because the RIS has a higher raw aperture gain due to its larger surface, while the SIM has an advantage in inter-user interference separation. When active beamforming is performed at the base station, the digital precoder already mitigates inter-user interference, making the advantage of the SIM less significant. As a result, the overall performance is dominated by the raw aperture gain.

\begin{figure}
    \centering
    \includegraphics[width=0.98\linewidth]{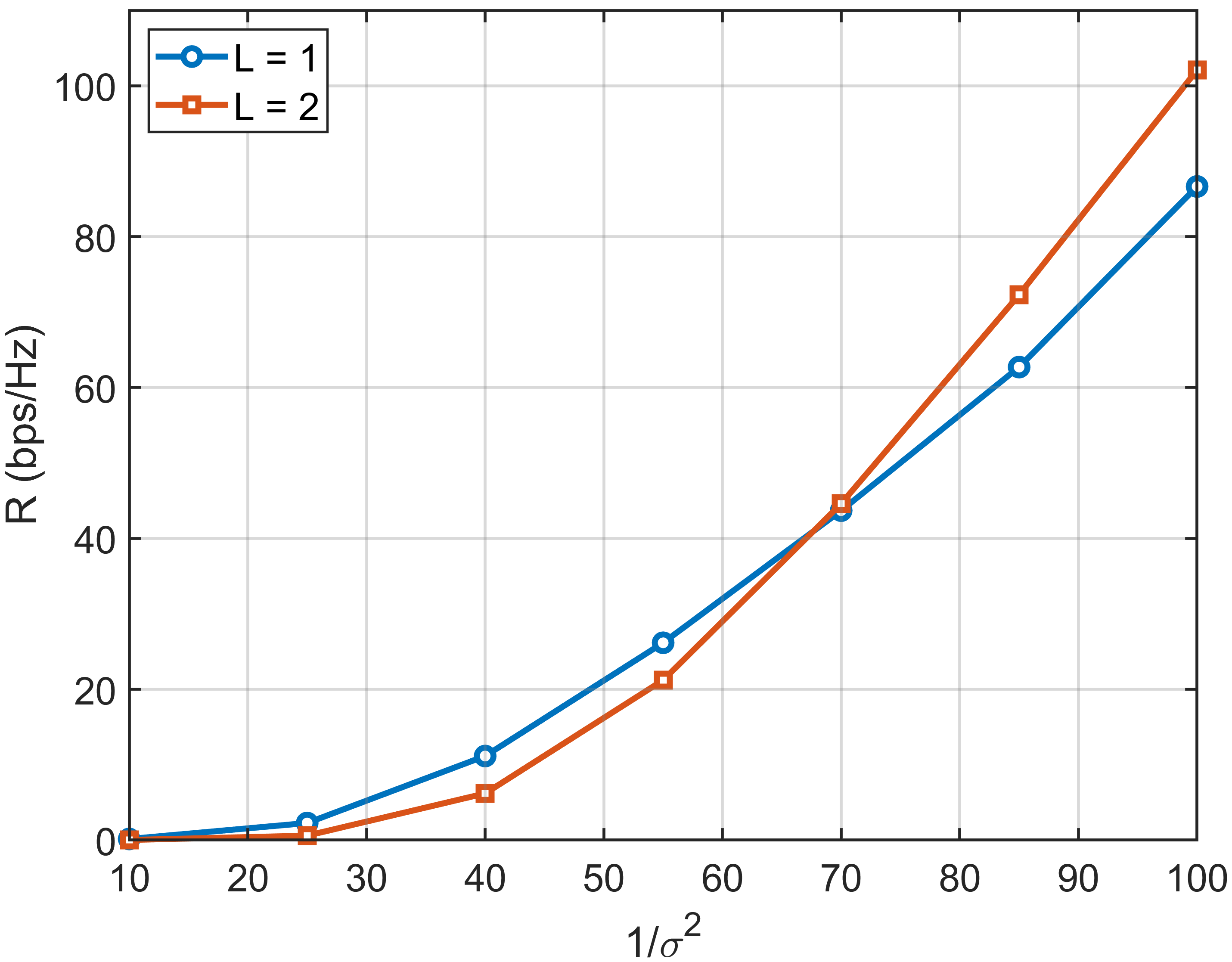}
    \caption{Variation of spectral efficiency with noise variance for $K = 6$ and $M_{total} = 100$}
    \label{fig:10}
\end{figure}

Next, we analyze the effect of the SNR (varying noise variance) on the spectral efficiency of both the RIS and the SIM in a multi-user system. Fig.~\ref{fig:10} presents a comparison between the two configurations for a total of $100$ antenna elements. At low SNR values (high noise variance), the RIS shows slightly better performance since noise dominates and the impact of inter-user interference remains limited. However, as the SNR increases, the SIM begins to outperform the RIS. This is because, in high-SNR regimes, inter-user interference becomes the primary performance-limiting factor, and the additional spatial degrees of freedom offered by the SIM enable more effective interference suppression and beam control. Consequently, the SIM achieves superior spectral efficiency at high SNR levels.

\begin{figure}
    \centering
    \includegraphics[width=0.98\linewidth]{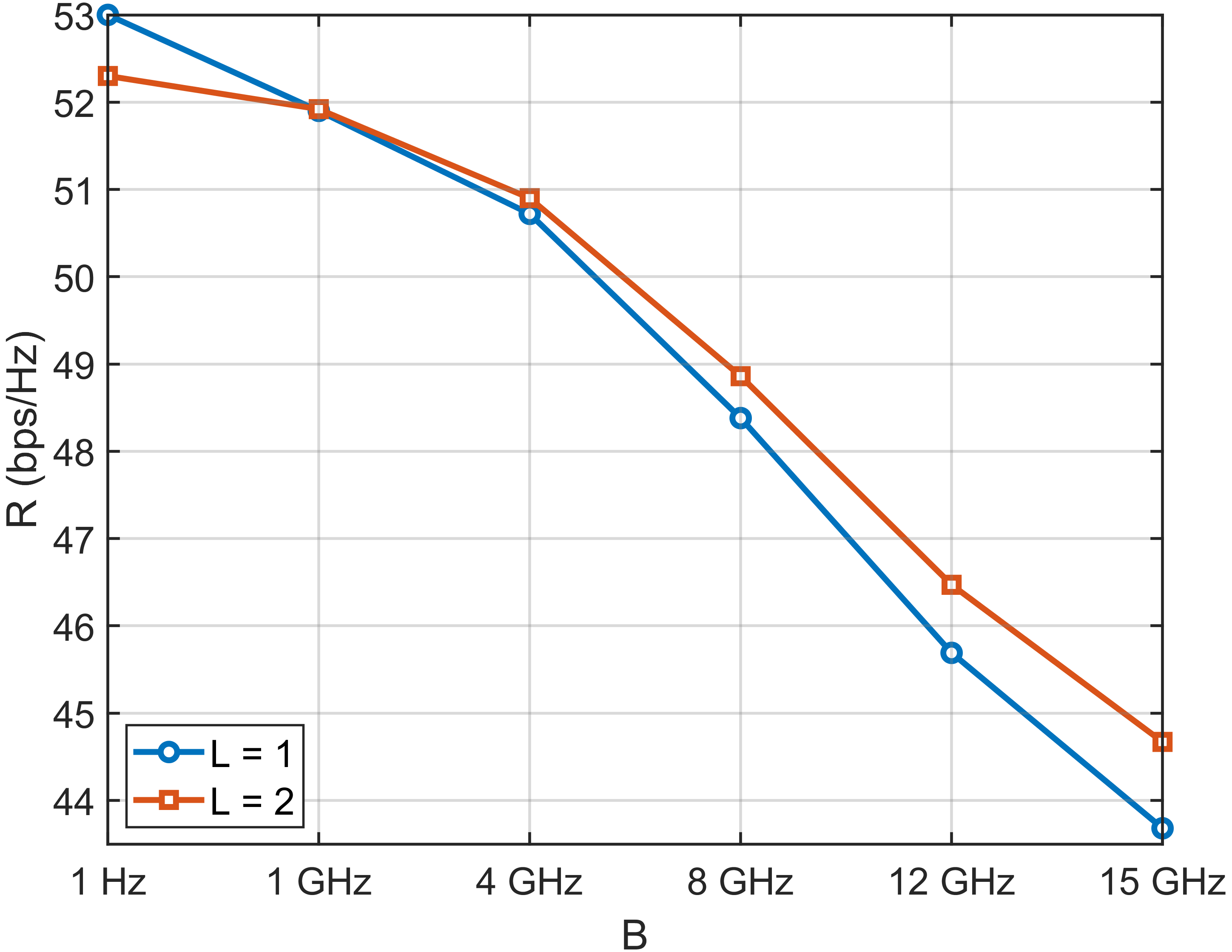}
    \caption{Variation of spectral efficiency with bandwidth $B$ for $K = 6$}
    \label{fig:17}
\end{figure}

In Fig.~\ref{fig:17}, we examine the impact of bandwidth on the spectral efficiency. It can be observed that in the narrowband regime, the single-layer configuration achieves slightly higher spectral efficiency. However, as the bandwidth increases, the multi-layer SIM begins to outperform the single-layer RIS, demonstrating its superior capability in mitigating frequency-selective effects and maintaining performance over wideband channels.

\begin{figure}
    \centering
    \includegraphics[width=0.98\linewidth]{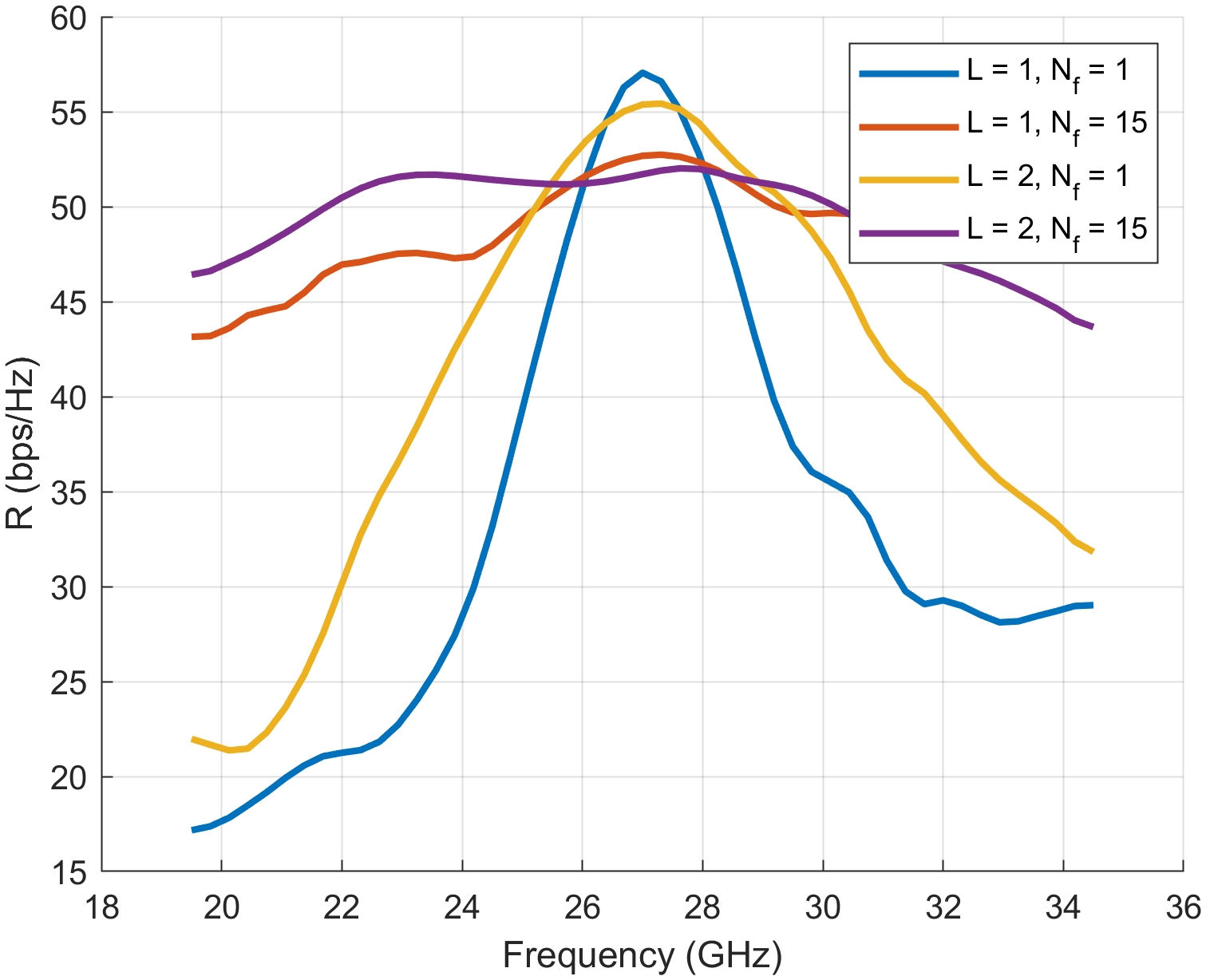}
    \caption{Variation of spectral efficiency with frequency}
    \label{fig:12}
\end{figure}

In Fig.~\ref{fig:12}, we illustrate the variation of spectral efficiency with frequency for both the RIS and the SIM when optimized over a single subband and multiple subbands, for a total of $K=5$ users and $144$ antenna elements. When optimized over a single subband ($N_f=1$) at the center frequency $f_c=27~\mathrm{GHz}$, the system achieves its highest gain near the center frequency but exhibits performance degradation at other frequencies due to frequency selectivity. The SIM shows a slightly lower peak gain at the center frequency compared to the single-layer RIS. However, it maintains better performance across the wider bandwidth owing to its multilayer structure, which provides additional spatial degrees of freedom. When the optimization is extended across multiple subbands ($N_f=15$), the overall gain at the center frequency slightly decreases, but the response becomes significantly flatter across the spectrum, resulting in improved robustness and more consistent spectral efficiency over the entire wideband range.

\begin{figure}
    \centering
    \includegraphics[width=0.98\linewidth]{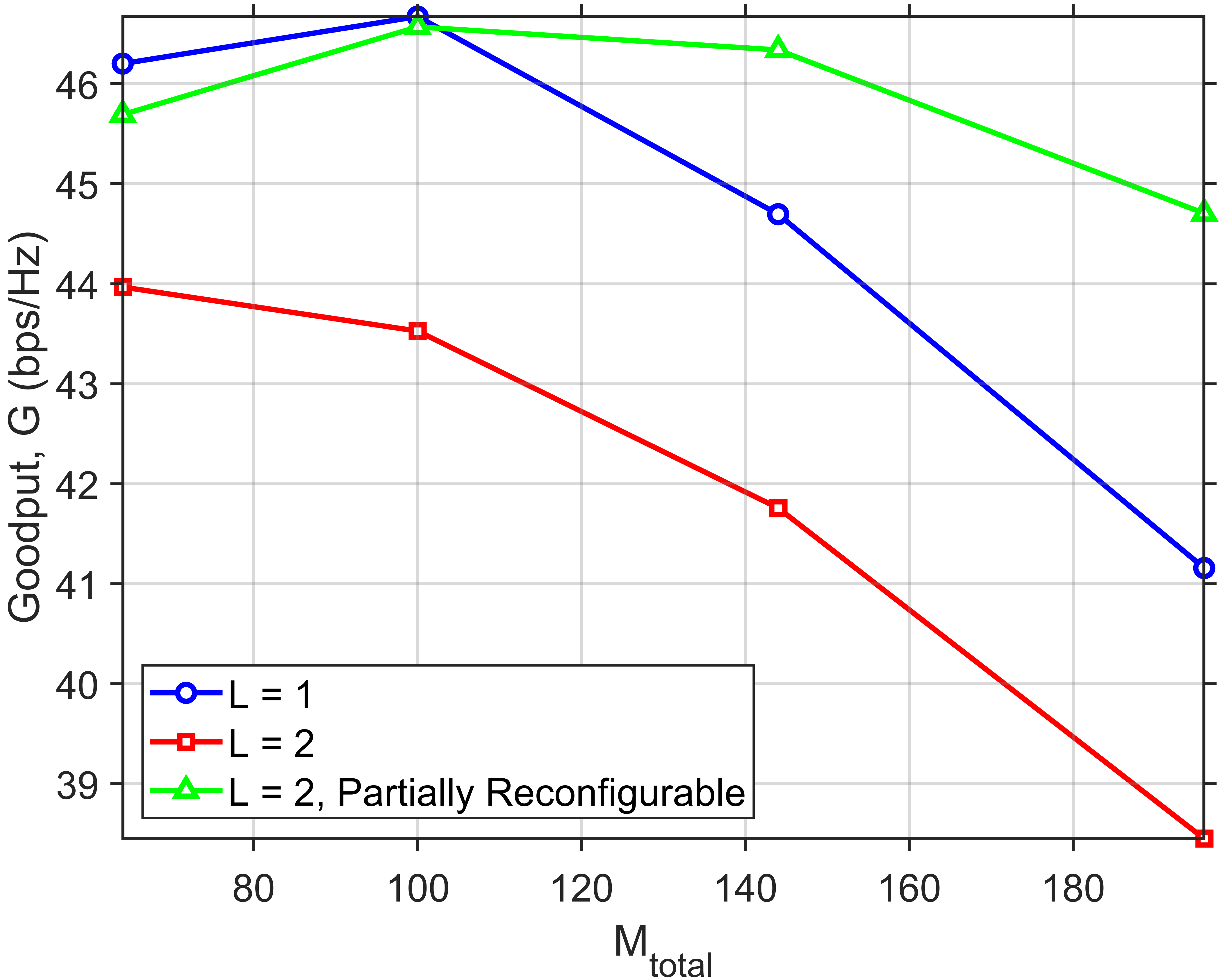}
    \caption{Variation of goodput with the number of antenna elements}
    \label{fig:13}
\end{figure}

Previously, we observed that the single-layer RIS performs better for a small number of users since inter-user interference is limited, whereas the SIM begins to outperform the RIS in higher SNR regimes and scenarios with more users, where interference suppression becomes more critical. In Fig.~\ref{fig:13}, we present the variation of \emph{goodput} with respect to the number of antenna elements for $K=2$ at $\mathrm{SNR}=110~\mathrm{dB}$. The results show that the SIM architecture provides additional flexibility through its multilayer design. Specifically, when only the second layer of the SIM is partially reconfigurable while the first layer remains fixed, the system achieves higher goodput compared to both the single-layer RIS and the fully reconfigurable SIM. This indicates that partial reconfiguration can offer a favorable trade-off between performance and complexity for systems with a small number of users, whereas a fully reconfigurable SIM is more advantageous in multi-user or high-SNR scenarios, where enhanced degrees of freedom are required to manage interference effectively.

\begin{figure}
    \centering
    \includegraphics[width=0.98\linewidth]{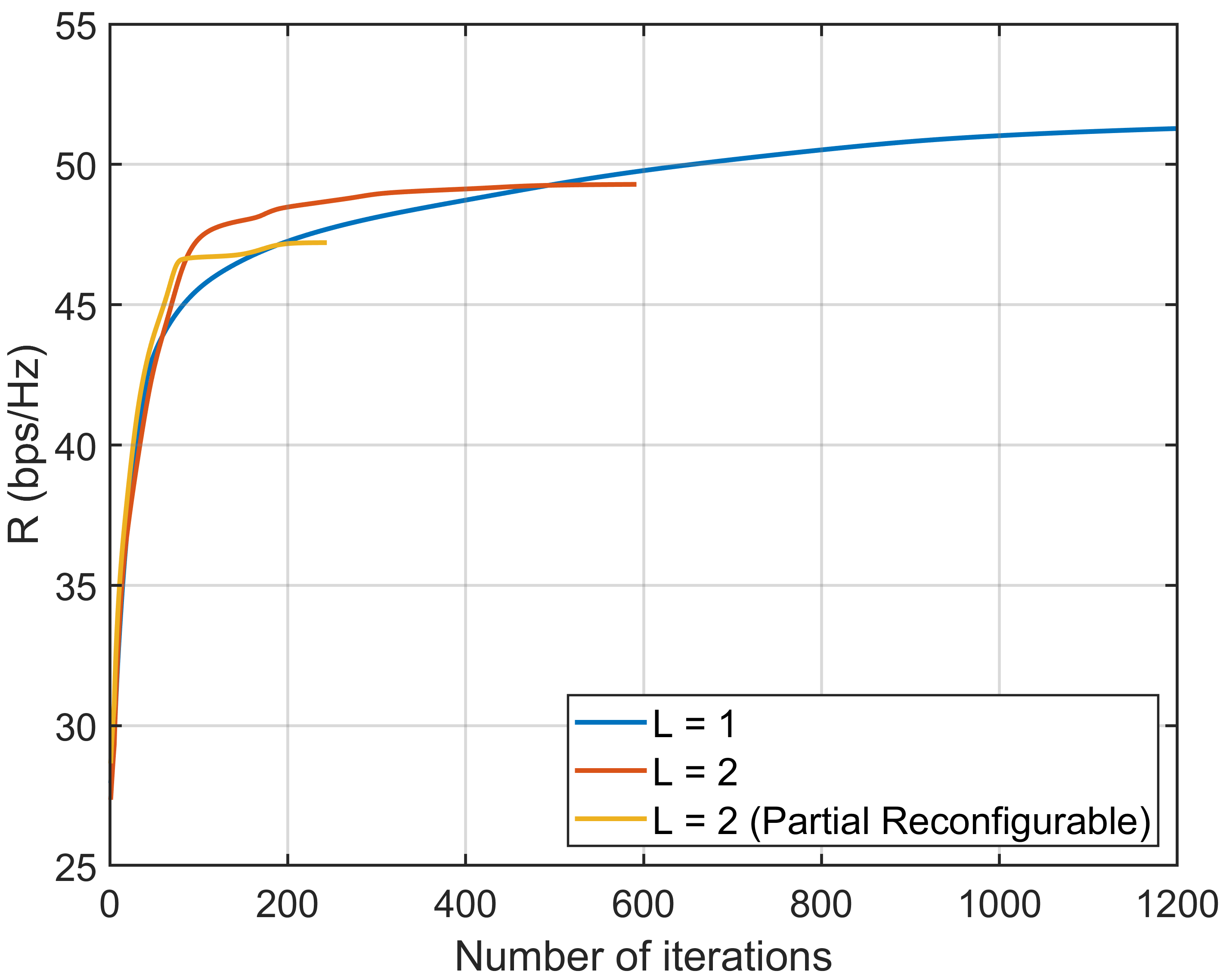}
    \caption{Convergence of gradient ascent algorithm}
    \label{fig:conv}
\end{figure}

{Finally, Fig.~\ref{fig:conv} shows the convergence of the proposed algorithm for $M_{\text{total}}=100$ and $K=4$. All schemes converge stably, and the SIM converges faster than the single-layer RIS. Moreover, the partially reconfigurable SIM not only uses fewer control bits but also converges within a few hundred iterations. In contrast, the single-layer RIS attains a slightly higher final spectral efficiency but requires more iterations, highlighting a trade-off between performance and complexity.}

\section{Conclusion}
We have presented a framework for a stacked intelligent metasurface (SIM)-assisted wideband downlink multiuser MIMO system that integrates carrier aggregation and phase optimization to enhance spectral efficiency. Using circuit theory, the proposed approach accurately captures mutual coupling, inter-layer reflections, and frequency selectivity, bridging electromagnetic behavior and system-level design. A joint optimization problem for SIM phase configuration and base station beamforming has been formulated, enabling adaptive carrier aggregation across favorable subbands. Moreover, a partially reconfigurable SIM has been proposed to reduce control overhead while maintaining system performance. We also have performed a comparative analysis of single-layer and multilayer structures with the same total number of antenna elements. We have observed that multilayer SIM achieves higher spectral efficiency than single-layer RIS, particularly in wideband, high-SNR, and high-user-density environments. However, when active beamforming is performed at the BS, the single-layer RIS starts to perform better. We have showed that incorporating mutual coupling improves spectral efficiency. Moreover, we have demonstrated that a partially reconfigurable SIM offers near-optimal performance with reduced complexity compared to fully reconfigurable metasurfaces.

\section*{Appendix —  Calculation of Gradients}

We present the derivation of the gradient with respect to the RIS phase parameters $\boldsymbol{\phi}$.  
Let
\begin{equation}
\begin{aligned}
\mathbf{G}(\boldsymbol{\phi})
&= \big(\mathbf{Z}_{SS}+\mathbf{Z}_{S}(\boldsymbol{\phi})\big)^{-1}, \\
\mathbf{G} &= \big[\mathbf{G}_{r,c}\big]_{r,c=1}^{2L},
\end{aligned}
\end{equation}
where $\mathbf{G}$ is partitioned into $2L\times 2L$ blocks $\mathbf{G}_{r,c}\in\mathbb{C}^{M\times M}$.
Then the end-to-end transfer matrix can be expressed as:
\begin{equation}
\mathbf{H}_Z
=\mathbf{Z}_{RS}\,\big(\mathbf{Z}_{SS}+\mathbf{Z}_S(\boldsymbol{\phi})\big)^{-1}\mathbf{Z}_{ST}
=\mathbf{Z}_{RS}\,\mathbf{G}(\boldsymbol{\phi})\,\mathbf{Z}_{ST}.
\end{equation}

The transmitter is connected to the first-facing ports of the first SIM layer, and the receiver is connected to the last-facing ports of the last layer.  
Let $\mathbf{S}_{\text{in}}\in\mathbb{R}^{(2LM)\times M}$ and $\mathbf{S}_{\text{out}}\in\mathbb{R}^{(2LM)\times M}$ denote the input and output selection matrices, respectively.  
Then,
\begin{equation}
\mathbf{G}_{2L,1} = \mathbf{S}_{\text{out}}^{\!\top}\,\mathbf{G}\,\mathbf{S}_{\text{in}}.
\end{equation}

To represent the effective impedance matrix, we define two auxiliary matrices, 
$\mathbf{H}_1 \in \mathbb{C}^{M\times N_t}$, corresponding to the first $M$ rows of $\mathbf{Z}_{ST}$, 
and $\mathbf{H}_2 \in \mathbb{C}^{K\times M}$, corresponding to the last $M$ columns of $\mathbf{Z}_{RS}$.

Thus, the effective baseband channel becomes
\begin{equation}
\mathbf{H}_{Z} = \mathbf{H}_2\,\mathbf{G}_{2L,1}\,\mathbf{H}_1.
\end{equation}
\subsection*{A. Gradient of the Local Impedance Matrix}
The gradient of local impedance block \eqref{eq:Z_matrix} characterized by $\phi_{\ell,m}$, is given by
\begin{equation}
\label{eq:F-local}
\frac{\partial \mathbf{Z}^{(\ell)}_{m}}{\partial \phi_{\ell,m}}
= -\,j Z_0
\begin{bmatrix}
1 + \left(\dfrac{\cos\phi_{\ell,m}}{\sin\phi_{\ell,m}}\right)^{\!2} &
\dfrac{\cos\phi_{\ell,m}}{\sin^{2}\!\phi_{\ell,m}} \\[8pt]
\dfrac{\cos\phi_{\ell,m}}{\sin^{2}\!\phi_{\ell,m}} &
1 + \left(\dfrac{\cos\phi_{\ell,m}}{\sin\phi_{\ell,m}}\right)^{\!2}
\end{bmatrix}.
\end{equation}

We denote these entries explicitly as
\[
\begin{bmatrix}
d_{11} & d_{12}\\[2pt]
d_{21} & d_{22}
\end{bmatrix}
\triangleq
\frac{\partial \mathbf{Z}^{(\ell)}_{m}}{\partial \phi_{\ell,m}}.
\]
Embedding \eqref{eq:F-local} within the global impedance matrix $\mathbf{Z}_S$ yields the sparse matrix $\mathbf{E}_{\ell,m} = \partial \mathbf{Z}_S / \partial \phi_{\ell,m}$, which is nonzero only at the ports corresponding to antenna element $(\ell,m)$.
\subsection*{B. Gradient for SIM}

The gradient of $\mathbf{G}(\boldsymbol{\phi})$ with respect to $\boldsymbol{\phi}$ follows from the matrix inverse differentiation property:
\begin{equation}
\label{eq:phi-G}
\frac{\partial \mathbf{G}}{\partial \phi_{\ell,m}}
= -\,\mathbf{G}\,
\mathbf{E}_{\ell,m}\,
\mathbf{G}.
\end{equation}
Substituting \eqref{eq:phi-G} into the derivative of the end-to-end transfer matrix gives
\begin{equation}
\frac{\partial \mathbf{H}_{Z}}{\partial \phi_{\ell,m}}
= -\,\mathbf{H}_2\,\mathbf{S}_{\text{out}}^{\!\top}\mathbf{G}\mathbf{E}_{\ell,m}\mathbf{G}\mathbf{S}_{\text{in}}\,\mathbf{H}_1.
\end{equation}
For a compact representation, let $\mathbf{e}_m\in\mathbb{R}^{M}$ be a vector whose $m$-th entry is one and all other entries are zero. Using this notation, we define

\[
\begin{alignedat}{2}
\mathbf{c}_1 &= \mathbf{G}_{2L,\,2\ell-1}\mathbf{e}_m, &\quad
\mathbf{c}_2 &= \mathbf{G}_{2L,\,2\ell}\mathbf{e}_m,\\[2pt]
\mathbf{r}_1^{\top} &= \mathbf{e}_m^{\top}\mathbf{G}_{\,2\ell-1,\,1}, &\quad
\mathbf{r}_2^{\top} &= \mathbf{e}_m^{\top}\mathbf{G}_{\,2\ell,\,1}.
\end{alignedat}
\]

Following the formulation in \cite{3}, we can obtain $c_1$, $c_2$, $r_1$, and $r_2$ using iterative solution and the matrix product simplifies to
\begin{equation}
\label{eq:F-lm}
\mathbf{F}_{\ell,m}
=
\big(d_{11}\mathbf{c}_1+d_{21}\mathbf{c}_2\big)\mathbf{r}_1^{\top}
+\big(d_{12}\mathbf{c}_1+d_{22}\mathbf{c}_2\big)\mathbf{r}_2^{\top}.
\end{equation}
Substituting into the derivative of $\mathbf{H}_Z$ yields
\begin{equation}
\label{eq:dHZ-final-G}
\frac{\partial \mathbf{H}_{Z,i}}{\partial \phi_{\ell,m}}
= -\,\mathbf{H}_2\,\mathbf{F}_{\ell,m}\,\mathbf{H}_1.
\end{equation}
\subsection*{C. Gradient of the Objective Function}

For user $k$ on subband $i$, define
\[
\begin{aligned}
\mathbf{h}_{k,i} &= (\mathbf{e}_k^{\top}\mathbf{H}_2)\,\mathbf{G}_{2L,1}\,\mathbf{H}_1,\\
p_{k,i} &= \mathbf{w}_{k,i}\mathbf{w}_{k,i}^H,\\
\mathbf{C}_{i,k} &= \sigma_i^2\mathbf{I}_{N_t}+\sum_{k'\neq k} p_{k',i}\,\mathbf{h}_{k',i}^{H}\mathbf{h}_{k',i},\\
\mathbf{v}_{i,k} &= \mathbf{C}_{i,k}^{-1}\,\mathbf{h}_{k,i}^{H}.
\end{aligned}
\]
The whitened gain term is
\[
s_{k,i} = p_{k,i}\,\mathbf{h}_{k,i}\,\mathbf{C}_{i,k}^{-1}\,\mathbf{h}_{k,i}^{H}
= p_{k,i}\,\mathbf{h}_{k,i}\,\mathbf{v}_{i,k}.
\]

Since $\mathbf{C}_{i,k}$ depends on $\boldsymbol{\phi}$ via $\mathbf{h}_{k',i}$, we have
\[
\partial \mathbf{C}_{i,k}^{-1} = -\,\mathbf{C}_{i,k}^{-1}\,(\partial \mathbf{C}_{i,k})\,\mathbf{C}_{i,k}^{-1}.
\]
Using the product rule and $\partial \mathbf{h}_{j,i}/\partial\phi_{\ell,m}
= -(\mathbf{e}_j^{\top}\mathbf{H}_2)\,\mathbf{F}_{\ell,m}\,\mathbf{H}_1$, we obtain
\begin{equation}
\label{eq:dski_full}
\begin{aligned}
\frac{\partial s_{k,i}}{\partial \phi_{\ell,m}}
&= 2\,p_{k,i}\,\Re\!\big\{(\partial\mathbf{h}_{k,i}/\partial\phi_{\ell,m})\,\mathbf{v}_{i,k}\big\} \\
&\quad -\,p_{k,i}\,\mathbf{h}_{k,i}\,\mathbf{C}_{i,k}^{-1}
\big(\partial \mathbf{C}_{i,k}\big)
\mathbf{C}_{i,k}^{-1}\,\mathbf{h}_{k,i}^{H}.
\end{aligned}
\end{equation}

Because
\[
\partial \mathbf{C}_{i,k}
= \sum_{k'\neq k} p_{k',i}\Big[
(\partial\mathbf{h}_{k',i})^{H}\mathbf{h}_{k',i} + \mathbf{h}_{k',i}^{H}(\partial\mathbf{h}_{k',i})
\Big],
\]
we simplify the whitening term using the trace trick as
\begin{equation}
\label{eq:whiten_simplified}
\begin{aligned}
\mathbf{h}_{k,i}&\,\mathbf{C}_{i,k}^{-1}
\big(\partial \mathbf{C}_{i,k}\big)
\mathbf{C}_{i,k}^{-1}\,\mathbf{h}_{k,i}^{H} \\
&= 2\sum_{k'\neq k} p_{k',i}\,
\Re\!\Big\{(\mathbf{h}_{k',i}\mathbf{v}_{i,k})^{*}\,
(\partial\mathbf{h}_{k',i}\,\mathbf{v}_{i,k})\Big\}.
\end{aligned}
\end{equation}

Therefore,
\begin{equation}
\label{eq:dski_compact}
\begin{aligned}
\frac{\partial s_{k,i}}{\partial \phi_{\ell,m}}
&= 2\,p_{k,i}\,
\Re\!\{\dot{\mathbf{h}}_{k,i}\mathbf{v}_{i,k}\} \\
&\quad -\,2\,p_{k,i}\!
\sum_{k'\neq k}\! p_{k',i}\,
\Re\!\{\!
(\mathbf{h}_{k',i}\mathbf{v}_{i,k})^{*}
(\dot{\mathbf{h}}_{k',i}\mathbf{v}_{i,k})\!\},
\end{aligned}
\end{equation}
\noindent
where $\dot{\mathbf{h}}_{k,i}\triangleq 
\partial\mathbf{h}_{k,i}/\partial\phi_{\ell,m}$. Hence, using Eq. \eqref{eq:gradJ_final_phi} and Eq. \eqref{eq:phi_update} we obtain the gradient of the objective function.


\bibliographystyle{IEEEtran}
\bibliography{paper_ref.bib}

\end{document}